\documentclass[aps,showpacs,superscriptaddress,amsmath,amssymb,preprint]{revtex4}
\usepackage{cmap}
\usepackage{graphicx}
\usepackage{epsfig}
\usepackage{verbatim}
\usepackage{amssymb}
\usepackage{amsmath}
\usepackage{xcolor}
\usepackage{comment}
\usepackage{ hyperref}
\usepackage[utf8x]{inputenc}
\newcommand{\bea}{\begin{eqnarray}}
\newcommand{\beq}{\begin{equation}}
\newcommand{\eea}{\end{eqnarray}}
\newcommand{\eeq}{\end{equation}}
\begin{document}
\title{Collective properties of a nuclear molecule composed of two heavy nuclei}
\title{Nuclear molecule of heavy nuclei}
\author{T. M. Shneidman}
\affiliation{Bogoliubov Laboratory of Theoretical Physics,
JINR, 141980 Dubna, Russia}
\author{R. G. Nazmitdinov}
\affiliation{Bogoliubov Laboratory of Theoretical Physics,
JINR, 141980 Dubna, Russia}
\affiliation{Dubna State University, 141982 Dubna, Russia}

\begin{abstract}
The model of a nuclear molecule that composed of two heavy nuclei is proposed.
To this aim the Hamiltonian of a dinuclear system is derived and
diagonalized in the basis of bipolar spherical functions. Analytical expressions, describing  excitations of highly deformed states of a nuclear molecule, are obtained.
A remarkable agreement between numerical and analytical results is demonstrated at the description of roto-vibrational excitations  in  $^{240}$Pu at low energies. We provide the prediction for the spectrum of hyperdeformed states of the nucleus $^{232}$Th, considering it as a nuclear molecule that consists of $^{132}$Sn+$^{100}$Zr nuclei. The angular distribution of fission fragments for the nucleus $^{240}$Pu have been analyzed as well and compared with available experimental data.
\end{abstract}
\maketitle

\section{Introduction}
\label{intro}

There is a wealth of experimental data pointing out on the usefulness of the nuclear molecule concept in their interpretation.
One of the first indications, confirming
the vitality of this concept, has been exhibited in the theoretical
description of alpha decay \cite{Bethe1936}.
Successful description of half-lives of $\alpha$-decaying nuclei,
assuming the formation of $\alpha$-particle on the surface of a heavy nucleus,
implies the formation of an asymmetric nuclear molecule (see, for example, \cite{Shneidman2003, Shneidman2011}).
Recently, the formation of $\alpha$-cluster configuration  was used to explain the reflection-asymmetric correlations in the vicinity of ground-state of actinides and rare-earth nuclei \cite{Shneidman2015}.
The existence of alpha-cluster molecules were also revealed in the knock-out and pickup reactions. To name just a few examples, we can point out
the recent $(p,p'\alpha)$ experiments on different $^{112-124}$Sn isotopes  \cite{Tanaka2021}; measurements of $\alpha$-cluster states
in various Ti isotopes by means of ($^{6}$Li,$d$) \cite{Yamaya1990},
($^{7}$Li,$t\alpha$) \cite{Fukada2009},
and ($^{3}$He,$d$)  \cite{Issatayev2023} reactions.
Another examples are high-spin resonant states,
observed at high excitation energies at $60\sim 70$ MeV in $^{28}$Si+$^{28}$Si, $^{24}$Mg+$^{28}$Si, and $^{24}$Mg+$^{24}$Mg reactions. Their properties can be explained by the formation of meta-stable states of the molecular type
\cite{Uegaki1993,Uegaki2012}.

This concept is quite suitable for interpretation
of deep inelastic and fusion-fission reactions.
In particular, it was proposed that the system of two fragments
(dinuclear system, or DNS)  bound together in the pocket of nucleus-nucleus potential is formed in the entrance channel  \cite{Volkov1982}.
The transfer of nucleons between fragments governs further course of the reaction.
Disintegration along the relative distance coordinate determines the mass and charge distributions of deep inelastic transfer and quasi-fission reactions.
The evolution along the mass asymmetry coordinate towards the compound nucleus
defines the cross section for a complete fusion (e.g., \cite{Antonenko}).
Similarly, in fission reactions, the scission configuration of
the fissile nucleus can be represented as a system of two fragments
located at the point of contact, i.e. alike the DNS \cite{Nix1965, Wilkins1976, Andreev2005,Rahmati2024}. The DNS decays is accompanied by modification of the fragment deformations and their angular orientation.  In fact, the collective motion, leading to the change in the orientation of the fragments, is responsible for generation of the angular momenta of the fission fragments (i.e.,\cite{Hess1984,H2,Misicu1999,Shneidman2002,Shneidman2003,AR2024,Randrup}).

At extreme nuclear elongations a nucleus can be treated as a molecular system as well \cite{Scheid}. In particular, the structure of highly deformed states of a harmonic oscillator  resembles  at certain particle numbers the molecular structure. In this case one fragment is spherical and the other is highly deformed, or both fragments are spherical magic nuclei \cite{Begtsson1981,Nazarewicz1992,ra}. The connection between highly deformed mirror-asymmetric states and cluster systems is also indicated in \cite{Aberg1994}. In fact, the calculations with the Woods-Saxon mean-field potential \cite{Cwiok1994} have shown that hyperdeformed (HD) states of heavy nuclei are presented as a molecular system with a good accuracy. For example, the HD state of $^{232}$Th can be described as a binary system that consists of $^{132}$Sn and $^{100}$Mo nuclei in the touching configuration. The energetic advantages of such a cluster system are related to the   fact that one of the clusters is a double  magic nucleus $^{132}$Sn, while the large quadrupole deformation of $^{100}$Mo allows a strong reduction of the Coulomb attraction.  Further, the HD state $^{180}$Hg can be described as a cluster system $^{90}$Zr+$^{90}$Zr \cite{Jonsson1997}.
Calculations in self-consistent mean field models demonstrate that under certain conditions the shell states are associated with a binary nuclear system \cite{Afanasjev2018}.

Development of new experimental facilities with an intense beam  of
monochromatic gamma-quanta \cite{ELINP,SAROV} opens a wide avenue in  investigation of dynamics of nuclear molecules and their states.
In particular, it would be possible to  study the spectroscopy of nuclei in strongly-elongated configurations. The excitation spectrum of such states  can be studied from the kinematic of reaction.

As mentioned above, attempts to formulate the nuclear molecular Hamiltonian were undertaken in \cite{Uegaki1993,Uegaki2012}.
The excitation spectrum of a system $^{26}$Mg+$^{26}$Mg has been investigated in terms of a few degrees of freedom, related to the relative angular vibrations of clusters. These clusters constitute the molecular system that rotates  as a whole.
The interaction between vibrational and rotational modes has been, however, neglected.
In fact, the Hamiltonian of the total kinetic energy with the Coriolis term in Ref.\cite{Uegaki2012} was replaced by an asymmetric rotator Hamiltonian, which does not seem to be entirely consistent. Consequently, it results in the underestimated value of the rotational moment of inertia. Such an approximation gives a minor influence on the analysis of properties of a molecular system that consists of light fragments.
Contrary, in case of heavy fragments (nuclei) this approximation may lead to incorrect conclusions regarding the evolution of fragments.

The major goal of the present paper is to derive a self-consistent  Hamiltonian
of a nuclear molecule, that enables to describe collective dynamics of a di-nuclear system when heavy nuclei are considered.
To this aim we use the Euler angles of our di-nuclear system as dynamical variables. The vitality and the validity of our approach will be demonstrated at the analysis of the experimental data related to the following situation:
one of the fragment of a nuclear molecule is spherical, while the other one  is an axially-symmetric quadrupole deformed object.
The interaction between the internal vibrations and rotation of the considered system will be treated exactly. A special attention will
be paid to the case when one of the fragments is strongly deformed. In this case we will present the analytical solutions that provide a basis for classification
of various rotational and vibrational modes in a nuclear molecule.

\section{Model}
\label{sec:model}

Let us assume that the compound nucleus ($A,Z$) can be considered as a molecular system,  consisting of an axially-symmetric quadrupole-deformed fragment ($A_1,Z_1$) and spherical fragment ($A_2,Z_2$), with $A_1+A_2=A$ and $Z_1+Z_2=Z$.
Hereafter, our discussion is restricted by the consideration of even-even fragments. Consequently, we use the equivalence between  the orbital and the total angular momentum of the deformed fragment.
For the sake of discussion the both fragments are characterized by a constant density. The distance between the centers of the fragments  is denoted by the magnitude of the vector ${\bf R}$, while the deformation of the fragment ($A_1,Z_1$) is defined by the quadrupole deformation parameter $\beta_{20}\equiv \beta_2$.
\begin{figure}[bth]
\vspace{-.1in}
\includegraphics[bb=0 1 256 237,scale=.75]{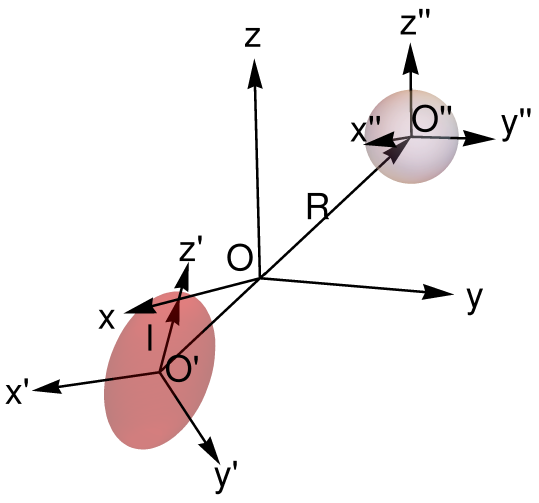}
\includegraphics[bb=0 0 251 262,scale=.75]{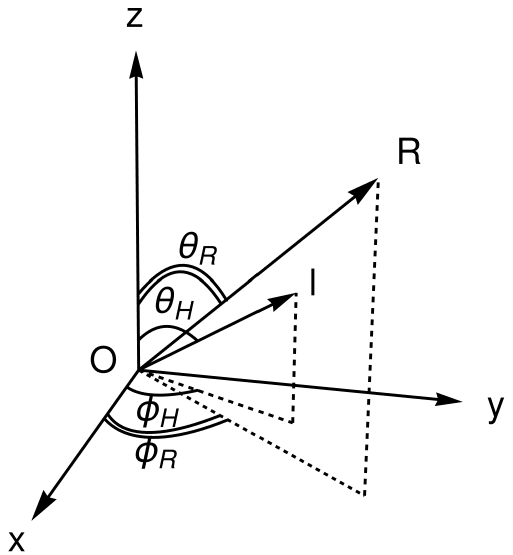}
\caption{(Color online) Left panel: a sketch of the molecular system. The vector ${\bf R}$ connects the centers of a quadrupole-deformed fragment and a spherical fragment. For the sake of illustration the magnitude of the vector ${\bf R}$  is much enhanced. The vector ${\bf l}$ coincides with the symmetry axis of a quadrupole-deformed fragment. The center of the laboratory system Oxyz coincides with the center of mass of the molecule.
Right panel: The orientation of the vector  ${\bf R}$ is defined by
the angles $\Omega_R=(\theta_R, \phi_R)$ with respect to
the laboratory system  $Oxyz$. The orientation of
the intrinsic coordinate system  $O'x'y'z'$ of the quadrupole-deformed
fragment  with respect to the laboratory
system is described by the angles  $\Omega_H=(\phi_H, \theta_H, 0).$
The axes of the intrinsic coordinate system $O"x"y"z"$ of
the spherical fragment are chosen to
be parallel with those of the laboratory system. }
\label{fig1}
\end{figure}

\begin{figure}[bth]
\begin{center}
\includegraphics[bb=0 1 221 188, scale=.75]{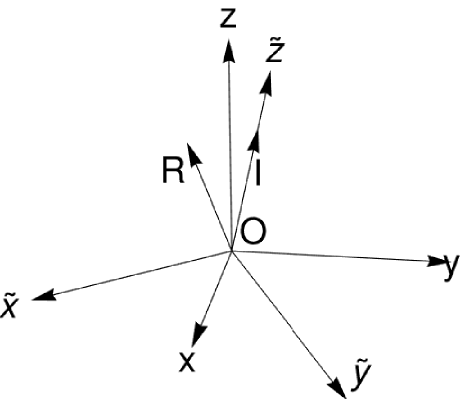}
\includegraphics[bb=0 1 220 175, scale=.75]{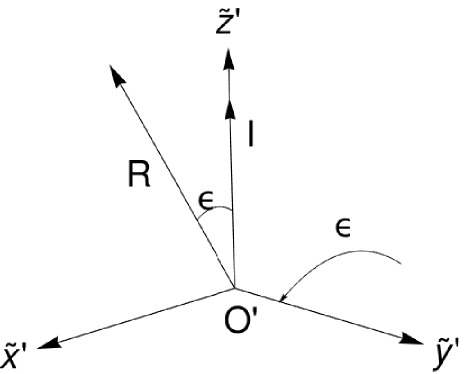}
\end{center}
\caption{Left panel: The coordinates of the body-fixed system
$O\tilde{x}\tilde{y} \tilde{z}$  with respect to the coordinates
of the laboratory system $Oxyz$. The axis $O\tilde{z}$  coincides
with the symmetry axis (the vector ${\bf l}$) of a deformed nucleus.
The vector ${\bf R}$ lies in the plane $O\tilde{x}\tilde{z}$.
Right panel: The coordinates of the body-fixed system  and a deformed nucleus with the axial symmetry. Both the vector ${\bf R}$ and the symmetry axis of the deformed nucleus lie in $O'{\tilde x}'{\tilde z}'$ plane.
}
\label{fig2}
\end{figure}

\subsection{Coordinate systems}
\label{sub1}

In order to define the collective motion in the molecule, we introduce the laboratory coordinate system $Oxyz$ (see Fig.\ref{fig1}) and the intrinsic coordinate systems of fragments. The center of the laboratory system coincides with the center of mass of the molecule. The spherical coordinates of the vector ${\bf R}$
in this system are denoted as $\{R,\theta_R,\phi_R\}$.

The center of the intrinsic coordinate system of the first (deformed) fragment $O'x'y'z'$ coincides with its center of mass.  The position of the center mass of the first fragment
with respect to the center of the molecule is defined by the vector
\begin{equation}{\bf R_1}=\frac{A_2}{A}{\bf R}
\label{R1}
\end{equation}
that connects  points  $O$ and $O'$. The axis $O'z'$  is chosen along the symmetry axis of the fragment. The orientation of the symmetry axis $O'z'$ with the respect of the laboratory system is defined by the Euler angles $\Omega_H=\{\phi_H,\theta_H,0\}$ .
Since the first fragment is assumed to be axially-symmetric we need only two angles to fix the orientation of the intrinsic coordinate system with respect to the laboratory system. The third Euler angle remains arbitrary and is chosen to be zero.

The center of the intrinsic coordinate system of the second (spherical) fragment $O''x''y''z''$ coincides with its center of mass. Since the second fragment
is assumed to be spherical, the axes of the system $O''x''y''z''$ are chosen to be parallel with those of the laboratory system.
Therefore, the translation from the laboratory system to the intrinsic coordinate system of the second fragment requires only the translation by the vector
\begin{equation}
{\bf R_2}=\frac{A_1}{A}{\bf R}.
\label{R2}
\end{equation}

Additionally, we introduce a body--fixed coordinate system (a molecular frame) $O\tilde{x}\tilde{y} \tilde{z}$ in such a way that
the direction of the axis $O\tilde{z}$  coincides  with the direction of the symmetry axis of the deformed fragment; and the direction of the vector ${\bf R}$ lies in the
plane $O\tilde{x}\tilde{z}$ (see Fig.\ref{fig2}).

Rotation from the laboratory system to the body--fixed coordinate system is defined by the Euler angles:
    \begin{eqnarray}
         \tilde{\Omega}=\{\theta_1=\phi_H,\theta_2=\theta_H,\theta_3=\phi_R \}.
    \label{bodyfixed}
    \end{eqnarray}
The orientation of the vector ${\bf R}$ in the body--fixed system is given by the angles $(\epsilon,0)$, and the cyclic components of this vector can be expressed as
\begin{eqnarray}
\label{rbfs}
   R_m=\sqrt{\frac{4\pi}{3}} R\  Y_{1 m}(\epsilon,0), \quad (m=-1,0,1).
    \label{rotation_1}
 \end{eqnarray}

\subsection{Kinetic energy}
\label{Sec_ke}

We start with the classical expression for the kinetic energy
\beq
\label{k1}
T = \frac{1}{2} \int \rho({\bf r}) \dot{{\bf r }}^2 d {\bf r }
 = \frac{1}{2} \Bigg\{\int \rho_1({\bf r}) \dot{{\bf r }}^2 d {\bf r } +\int \rho_2({\bf r}) \dot{{\bf r }}^2 d {\bf r }\Bigg\}\,,
\eeq
where   a weak overlap is assumed between the fragments :
$\rho({\bf r})=\rho_1({\bf r})+\rho_2({\bf r})$;
and $\rho_i({\bf r})$, ($i=1,2$) are the fragment densities.
It is convenient for each integral to perform a parallel translation to the center of the corresponding fragment.
\begin{eqnarray}
\label{ke4}
T&=&\frac{1}{2} \int \rho_1({\bf r}) ({\bf \dot{r}_1+\dot{R}_1})^2 d {\bf r_1 }+\frac{1}{2} \int \rho_2({\bf r}) ({\bf \dot{r}_2+\dot{R}_2})^2 d {\bf r_2} \nonumber \\
&=& \frac{1}{2} \int \rho_1({\bf r_1}) {\bf \dot{r}_1}^2 d {\bf r_1 }+
\frac{1}{2} \int \rho_2({\bf r_2}) {\bf \dot{r}_2}^2 d {\bf r_2 }\\
&+&  {\bf \dot{R}_1}\int \rho_1({\bf r_1}) {\bf \dot{r}_1} d {\bf r_1 }{\bf +\dot{R}_2}\int \rho_2({\bf r_2}) {\bf \dot{r}_2} d {\bf r_2 }\nonumber \\
&+&\frac{1}{2}A_1 {\bf \dot{R}_1}^2+\frac{1}{2}A_2 {\bf \dot{R}_2}^2\,.\nonumber
\end{eqnarray}
Assuming a constant density inside fragments and that vectors ${\bf r_1}$ and ${\bf r_2}$ are measured from the center of mass of the corresponding fragment, we have
\begin{eqnarray}
 \int_{V_i} \rho_i({\bf r}) {\bf {\dot r}_i } d {\bf r_i }=
 \frac{d}{dt} \int_{V_i} \rho_i({\bf r}) {\bf  r_i } d {\bf r_i }=0, \quad (i=1,2)\,.
\label{ke5}
\end{eqnarray}
We recall that the second fragment  is considered to be spherical, i.e., it has no rotational excitations. As a result, in the spherical coordinates
we obtain:
\begin{eqnarray}
    {\bf \dot{r}_2}^2=\frac{4\pi}{3}{\dot r}^2 \sum_{\mu=-1,0,1} |Y_{1\mu}(\theta_2,\phi_2)|^2={\dot r}^2.
\end{eqnarray}
Thus, the part of the kinetic energy, related to the second fragment,
may be associated with dynamic density fluctuations. Excitations, related to these
processes, are located in the region of giant resonances (GRs). In turn,
our analysis is focused on the region of low-lying excitations,
where contributions of GR tails can be neglected safely (e.g., \cite{sev}).
Therefore, it is reasonable to suppose that
\begin{equation}
\label{rr2}
\int_{V_2}\rho_2({\bf r_2}) {\bf \dot{r}_2}^2 d {\bf r_2 } \approx   0.
\end{equation}
In virtue of Eqs.(\ref{R1}), (\ref{R2}), (\ref{ke5}), (\ref{rr2}), the kinetic energy
(\ref{ke4}) transforms to the form:
\begin{eqnarray}
\label{ke5a}
T& = &\frac{1}{2} \int_{V_1} \rho_1({\bf r_1}) {\dot {\bf{r}}_1^2} d {\bf r_1 }+
\frac{1}{2} \mu {\bf {\dot{R}}^2 \,.}
\end{eqnarray}

First, we evaluate the second term. After a few algebraic manipulations (see Appendix \ref{ap1})
we  obtain the expression:
\begin{eqnarray}
T=\frac{1}{2} \int_{V_1} \rho_1({\bf r_1}) {\bf \dot{r}_1^2} d {\bf r_1 } +
\frac{1}{2}\mu \dot{R}^2 +
\frac{1}{2}\mu R^2 (\dot{\theta}^2_R+\sin^2 \theta_R \dot{\phi}_R^2),
\label{ke6}
\end{eqnarray}
where $\mu=m_0 A_1 A_2/{A}$ is the reduced mass of the molecule and $m_0$
is the nucleon mass.

Second, we evaluate the kinetic energy of the first fragment.
Using the results from Appendices
\ref{ap1}, \ref{dfrag},
we  obtain finally the expression for the classical kinetic energy in the laboratory frame:
\begin{eqnarray}
T=\frac{1}{2}\Im_H (\dot{\theta}^2_H+\sin^2 \theta_H \dot{\phi}_H^2)+\frac{1}{2}\mu \dot{R}^2+\frac{1}{2}\Im_R (\dot{\theta}^2_R+\sin^2 \theta_R \dot{\phi}_R^2),
\label{ke7}
\end{eqnarray}
where $\Im_H$ is the moment of inertia of the fragment $(A_1,Z_1)$, and $\Im_R=\mu R^2$.
We rewrite expression (\ref{ke7}) in the compact form
\begin{equation}
T=\frac{1}{2}\sum_{i, j =1}^5 G_{i j} \dot{\theta}_i\dot{\theta}_j\,.
\end{equation}
Here $\theta_1=\theta_H, \theta_2=\phi_H, \theta_3=\theta_R, \theta_4=\phi_R, \theta_5=R$. The diagonal matrix elements of the metric tensor
$G_{ij}$ are given by
$G_{11}=\Im_H$, $G_{22}=\sin^2{\theta_H}$, $G_{33}=\Im_R$,
$G_{44}=\sin^2{\theta_R}$, $G_{55}=\mu$; all nondiagonal elements are zero.
By means of the Pauli quantization procedure \cite{Pauli}
\begin{eqnarray}
\label{ap4}
&&\hat T=-\frac{\hbar^2}{2}\sum_{i j } \frac{1}{\sqrt{G}}\frac{\partial }{\partial \theta_i}\sqrt{G}(G^{-1})_{i j}\frac{\partial }{\partial \theta_j}\,,
\end{eqnarray}
we obtain the operator of the kinetic energy :
\begin{eqnarray}
&&  \hat T= - \frac{\hbar^2}{2\mu R^2} \frac{\partial}{\partial R} R^2 \frac{\partial}{\partial R}+
\frac{\hbar^2{\hat L}_R^2}{2\Im_R} +\frac{\hbar^2{\hat L}^2_H}{2\Im_H} \,,
\label{kinen}
\end{eqnarray}
where the angular momentum operators are defined as:
 \begin{eqnarray}
&& {\hat L}^2_i=-\frac{1}{\sin
\theta_i}\frac{\partial}{\partial\theta_i}\sin \theta_i
\frac{\partial}{\partial\theta_i}-\frac{1}{\sin^2
{\theta_i}}\frac{\partial^2}{\partial \phi_i^2},
\nonumber \\
&&\hfill(i=R,H).
\label{enoper}
\end{eqnarray}

\subsection{Potential Energy}
\label{poten}

There are several methods to calculate the potential energy of interaction between the fragments of molecular system.
 To elucidate the physical mechanisms, underlying the potential
energy, it is useful  to analyze qualitatively the nature of the forces
that determine the potential energy.

\begin{figure}[bth]
\begin{center}
\includegraphics[bb=0 0 393 353,scale=.5]{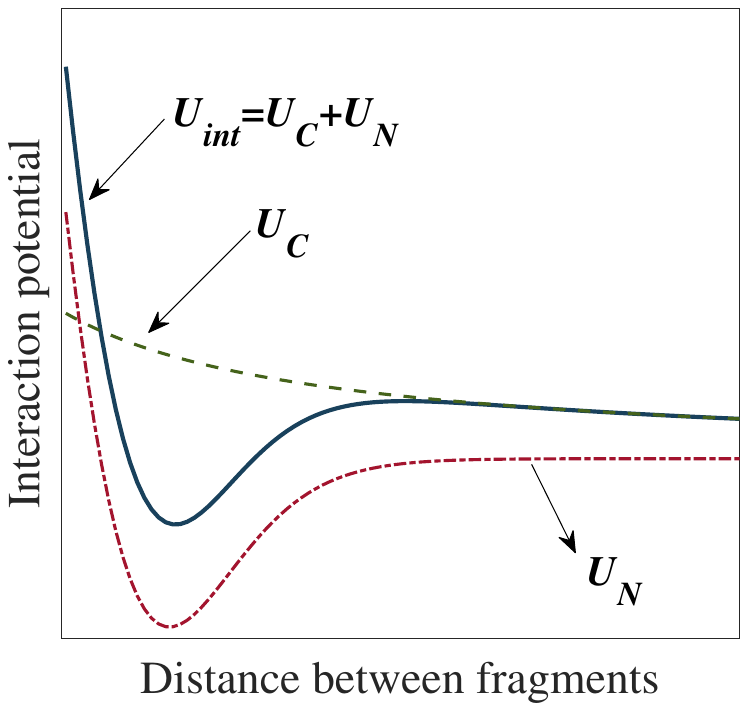}
\end{center}
\vspace{-.5cm} \caption{A sketch of the effective interaction potential between
fragments as a function of a relative distance $x$. Here $U_C$ is the Coulomb potential, $U_N$ is a nuclear interaction potential.}
\label{scheme}
\end{figure}

For the sake of discussion, we consider two spherical fragments.
The potential energy of the molecule $U=U_N+U_{C}$ is determined by a sum
of a nuclear $U_N$ and the Coulomb $U_C$ interaction potentials (see Fig.\ref{scheme}).
The Coulomb term can be written as
\begin{eqnarray}
\label{coul}
U_C(x)=\frac{e^2 Z_1 Z_2}{R_1+R_2+x},
\end{eqnarray}
where $x$ is the distance between the fragment's tips.

Suppose that a nuclear interaction between two fragments can be approximated by Migdal forces \cite{Migdal}, which assume different nucleon-nucleon interaction strengths in the interior and exterior of a compound system.
The nuclear potential between two fragments can be obtained by folding
the nucleon-nucleon interaction with fragment densities, which results in a molecular type potential \cite{Adamian1996}.
According to a general wisdom, such a potential resembles an effective potential of two interacting atoms in a molecule.
Consequently, we approximate it by the Morse type potential:
\begin{eqnarray}
\label{un}
U_N(x)=-F_0 \left [2\exp{(-x/b)} -\exp{(-2 x/b)} \right ].
\end{eqnarray}
Since the nuclear interaction is proportional
to the volume of overlap between two fragments,
evidently, the potential should diminish at the distances
larger than the sum of diffuseness regions of nuclei $x_d$.
Thus, considering the decrease
of the potential by  $e$ times, we obtain
\beq
\frac{U_N(x_d)}{U_N(0)}\approx e^{-1}\Rightarrow x_d=b=a_1+a_2\,,
\eeq
where $a_i$ ($i=1,2$) are the diffuseness parameters of the nuclear densities.

\begin{figure}[t]
\begin{center}
\includegraphics[scale=0.45]{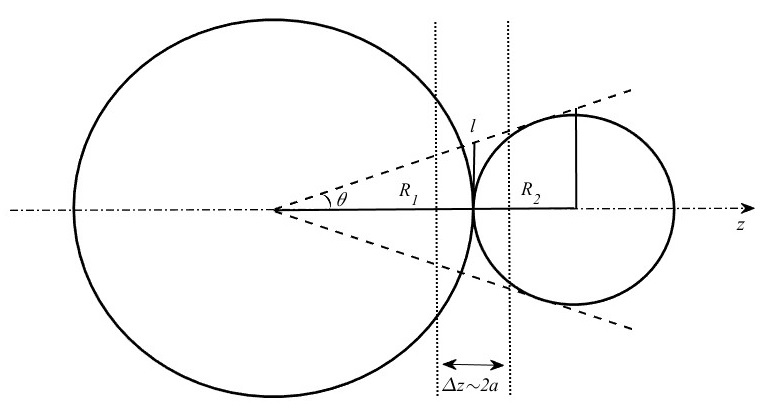}
\end{center}
\vspace{-.5cm} \caption{Scetch of the binary system at the touching configuration.}
\label{scheme_touching}
\end{figure}

Let us estimate the magnitude $F_0$ of a nuclear interaction potential
at the touching configuration. The volume of the
overlap is (see Fig.~\ref{scheme_touching})
\begin{eqnarray}
\delta V_{ov.} \approx \pi l^2 \Delta z,\quad  \Delta z \approx 2a, \quad
a =(a_1+a_2)/2\,.
\label{app1}
\end{eqnarray}
To estimate $l$, we draw a tangent line from the center of the
heavy fragment to the surface of
the light fragment (see Fig.~\ref{scheme_touching}).
Defining  $l$ as the length of the perpendicular from the tangent line to the vector ${\bf R}$ at $z=R_1$, we obtain
\begin{equation}
\label{line}
l=R_1\tan\theta=R_1\times \frac{R_2+\eta}{R_1+R_2}\approx \frac{R_1R_2}{R_1+R_2}\,.
\end{equation}
Here, $R_2+\eta$ is the perpendicular, connecting the continuation
of the tangent line with the center
of the light fragment. The  approximation (\ref{line}) is well justified, since $\eta\ll (R_1+R_2)$.
Using Eqs.(\ref{app1}--\ref{line}), we estimate
the nuclear interaction (\ref{un}) at the touching point as
\begin{eqnarray}
    U_N(x=0)=-F_0 \approx -f_0 \pi a \left (\frac{R_1 R_2}{R_1+R_2}\right )^2,
    \label{app5}
\end{eqnarray}
where $f_0$ is a fitting parameter of the nucleus-nucleus interaction. The expression (\ref{app5}) has the form of  a proximity potential \cite{Blocki1977} used to describe the nuclear interactions between two nuclei in the vicinity of touching (see also Ref.\cite{Proximity}).

Our aim is to find the extremum values of the effective potential
$U(x)$.
There is a point between the minimum and maximum of the potential $U$ where it varies
most strongly.
Since at these distances, the nuclear part of the interaction changes much faster than the Coulomb interaction, we can neglect the change of the latter one and conclude that $U_N^{''}(x=x_0)=0$ at the above point.
Solving this equation, we obtain
\bea
x_0=b \ln{(2)} \approx 0.693 b.
\eea
Expanding the first derivative of the potential $U_N$ around this point in the Taylor series up to the second order, we obtain
\begin{eqnarray}
U^{'}_N(x) \approx \frac{F_0}{2 b}-\frac{F_0}{2 b^3}(x-x_0)^2.
\end{eqnarray}
Keeping in mind a slow variation of the Coulomb contribution, we  find
\begin{eqnarray}
U_C^{'}(x)=-\frac{U_C(x)}{R_1+R_2+x} \approx -\frac{U_C(x_0)}{R(x_0)}\,, \quad
R(x_0)=R_1+R_2+x_0\,.
\end{eqnarray}
The extremum  condition for the interaction potential
$U_C^{'}(x)+U^{'}_N(x)=0$ yields
\begin{eqnarray}
    x_{1,2}& = &x_0\pm b \sqrt{1-\frac{U_C(x_0)}{R(x_0)}\times{\frac{2b}{F_0}}}\,.
\end{eqnarray}
The pocket in the interaction potential persists  if the condition
\begin{eqnarray}
\label{ulim}
    1- \frac{U_C(x_0)}{R(x_0)} \frac{2 b}{F_0}\geqslant 0
\end{eqnarray}
is met.  To obtain the upper limit of this inequality,
we use the following parameters: $b=2a$, where $a \approx 0.55$ fm is the diffuseness parameter
of nuclear densities; $f_{0}\approx 3.3$ MeV/fm$^3$;
$R_i=1.2 A^{1/3}$ fm, ($i=1,2$) are equivalent spherical radii of interacting nuclei.  As a result, we obtain
the following condition
\begin{eqnarray}
\label{con22}
\frac{Z_1 Z_2}{(R_1+R_2+0.762)^2} \left (\frac{R_1+R_2}{R_1 R_2} \right )^2 \leqslant 1.8\,
\end{eqnarray}
for the existence of a pocket in the molecule potential energy.
Evidently, the upper limit of this condition is reached at:
i)$Z_1 =Z_2=Z/2$; ii) $R_1=R_2=1.2 (A/2)^{1/3}$; where
$Z$ and $A$ are the charge and the mass of the molecule, respectively.
Consequently, Eq.~(\ref{con22}) yields the following result:
\beq
\label{prox}
Z^2/A^{4/3}\lesssim 6\,.
\eeq
 It is notable to mention that for the nuclear molecule that consists of two $^{232}$U nuclei ($Z=92)$ the above equation yields the value $\sim 6$ \cite{Zag2006}.
This equation defines the stability of a molecular system with respect to the decay due to the interplay between the Coulomb repulsion and nuclear interactions. It resembles the expression for the fissility parameter of the liquid drop, used to estimate its stability determined by the interplay between the surface and the Coulomb energies. We belief that Eq.(\ref{prox}) can be considered as a possible criterion for the stability of a nuclear molecule.

It appears that the barrier disappears for nearly symmetric molecules with $Z>100$, which  is characteristic  for  molecules of actinide and transactinide mass regions.
Our analysis will be focused on the properties of molecular systems in these regions. Thus, the interplay between the short-range nuclear attraction and the Coulomb repulsion reaches its local minimum at the distances close to the touching configuration of fragments. Furthermore, this result is valid for an arbitrary relative orientations of the fragments.

In general, the interaction potential between an axially-symmetric quadrupole deformed and spherical fragments of the
molecule can be expressed through a multipole expansion of the form:
\begin{eqnarray}
\label{gpot}
    &&U(R, \Omega_H,\Omega_R)=\sum_{l}U_{l}(R) \left [ Y_l(\theta_H,\phi_H) \times Y_l(\theta_R,\phi_R)\right ]_{(00)} \nonumber \\
    &=& \sum_{l} \frac{(-1)^l}{\sqrt{2l+1}}U_{l}(R)\left ( Y_l(\theta_H,\phi_H) \cdot Y_l(\theta_R,\phi_R)\right ) \nonumber \\
    &=&\sum_{l} \frac{(-1)^l \sqrt{2l+1}}{4\pi}U_{l}(R) P_l(\cos{\epsilon})=U(\beta_2, R,\epsilon)\,.
\end{eqnarray}
Here, the angle $\epsilon$ is defined on Fig.\ref{fig2}.
For an axially-symmetric system the summation runs  over even $l$ only.

\subsection{The Hamiltonian and eigenfunctions}
\label{eig}

The results, obtained in Secs.\ref{Sec_ke},\ref{poten}, yield the potential and kinetic energy operators for
the quantum Hamiltonian of the nuclear molecule.
Our aim is to choose a suitable basis and to solve the eigenvalue problem. However,  a few comments are in order.

To begin with, we suppose that vibrational  energies, associated with the excitations of the relative distance coordinate $R$,
are large in comparison with rotational energies. The latter are generated by the rotation of the heavy fragment and  by the rotation of the molecule itself relative to the laboratory coordinate system
(hereafter, the relative rotation), i.e.,
\begin{eqnarray}
\hbar \omega_R \gg \hbar^2/\Im_H,  \hbar^2/\Im_R \,.
\end{eqnarray}
We recall that $\Im_H$ and $\Im_R$ are the moments of inertia of
the heavy fragment and the relative rotation, respectively.
Indeed,  approximating the radial part of the potential energy in the region of pocket by the oscillator, one can estimate the frequency  $\hbar \omega_R$. Calculations yield $\hbar \omega_R \approx 10$  MeV, which is sufficiently larger than the rotational energies in the region of angular momenta of interest $l \leqslant (10-12) \hbar$. Therefore, for the description of low-lying excitations of  a DNS we neglect excited states associated with $R$.

Taking into account the above facts, we employ the Born-Oppenheimer approximation.
Consequently, we consider the wave function  as a product of the wave functions of slow and fast coordinates ($\epsilon$ and R, respectively):
\begin{eqnarray} \label{B-O-approximation}
\Psi=\frac{\psi(R)}{R}\mathcal{Y}(\epsilon).
\end{eqnarray}
We recall that in this approximation the motion in the fast coordinate creates an additional potential in the description of the slow coordinate.
In the equation for the radial wave function $\psi(R)$,  the kinetic energy in the variable $\epsilon$ is neglected:
\begin{eqnarray}\label{B-O-radial}
\left [-\frac{\hbar^2}{2 \mu } \frac{\partial^2}{\partial R^2}+U(\beta_2,R,\epsilon)\right ] \psi(R)=E(\epsilon) \psi(R)\,,
\end{eqnarray}
where the energy $E(\epsilon)$ is
\begin{eqnarray}
\label{B-O-radial-E}
E(\epsilon)\approx\langle T_{R}\rangle+U(\beta_2,R_m(\epsilon),\epsilon)\,.
\end{eqnarray}
Here, $\langle T_{R}\rangle$ is the average kinetic energy.
Since the motion, related to the coordinate $R$, is fast comparing to
the angular motion (rotation), we have assumed that
for each value of $\epsilon$, the coordinate $R$ takes its average value
\begin{eqnarray}
    R \rightarrow R_m=R_m(\epsilon)\,.
\end{eqnarray}
This value is identified as  a touching distance between the fragments for the particular value of $\epsilon$.
Zero-point vibrations in the coordinate $R$ contribute to the potential energy of the Hamiltonian, describing the angular rotation.

Based on the above discussion, the potential (\ref{gpot}) reaches
its local minimum at the distances close to the touching configuration of fragments at any value of the angle $\epsilon$.
As a function of $\epsilon$, the potential energy is a symmetric function.  Once the spherical fragment is situated at one of the poles of the deformed fragment, the potential energy has a minimum.
If the spherical fragment is  situated in the equatorial plane,
 the potential energy has a maximum. Thus, it has a form of a double well with  two minima at $\epsilon=0$ and $\epsilon=\pi$, separated by the barrier at
$\epsilon=\pi/2$. The simplest function, possessing such a behavior, is well approximated by
$\sin^2\epsilon$.
Consequently,
the interaction potential (\ref{gpot}) can be presented in the following form
\bea
\label{potential}
&&U(\beta_2,R_m,\epsilon) \approx
U_0(\beta_2) + \frac{C(\beta_2,R_m)}{2} \sin^2{\epsilon}=\\
&=&\Big (U_0(\beta_2) + \frac{C(\beta_2,R_m)}{3}\Big )P_0(\cos{\epsilon}) -
\frac{C(\beta_2,R_m)}{3}P_2(\cos{\epsilon}) \nonumber\,.
\eea
The parameter $C$ can be obtained
within a microscopic approach or by fitting experimental data.
Below we will provide the approximate evaluation for this parameter at the special cases (see Sec.\ref{3a}).

Since the first term in Eq.(\ref{B-O-radial-E}) does
not depend on $\epsilon$, it can be omitted, and the equation
for the wave function $\mathcal{Y}(\epsilon)$ takes the form
 \begin{eqnarray}\label{B-O-angular}
  \left (\frac{\hbar^2 \hat{L}^2_{R}}{2 \mu R^2_m}+\frac{\hbar^2 \hat{L}^2_{H}}{2 \Im_H}
  + \frac{C(\beta_2,R_m)}{2} \sin{\epsilon}^2\right)\mathcal{Y}(\epsilon)
=(E-U_0)\mathcal{Y}(\epsilon)\,.
 \end{eqnarray}
 Here,  the expression (\ref{potential}) has been used for the definition
 of the potential energy, and we preserve the notation $\Im_R=\mu R^2_m(\epsilon)$. The term $U_0$ gives a constant shift, and, therefore, can be safely omitted.
As a result, taking into account Eq.(\ref{kinen}-\ref{enoper}),
Eq.({\ref{potential}}), we obtain
the following Hamiltonian for the angular motion in the nuclear molecule
\begin{eqnarray}
\label{fullH}
&&\hat H = \hat H_{rot}+V_{int}\,, \nonumber\\
\label{hamiltonian}
&&\hat H_{rot} = \frac{\hbar^2}{2 \Im_R}\hat L^2_R+\frac{\hbar^2}{2 \Im_H}\hat L^2_H\,,\\
&&V_{int} = \frac{C(\beta_2,R_m)}{3} \left (1 -\frac{4\pi}{\sqrt{5}} \left [Y_{2}(\theta_H,\phi_H) \times Y_{2}(\theta_R,\phi_R)\right ]_{(00)}\right )\nonumber\,.
\end{eqnarray}

The bipolar spherical functions
\begin{eqnarray}
&&| L_H L_R; I M\rangle \equiv \left [Y_{L_H}(\theta_H,\phi_H) \times Y_{L_R}(\theta_R,\phi_R)\right ]_{(IM)} \nonumber \\
& \equiv & \sum_{M_H, M_R}C^{IM}_{L_H M_H \ L_R M_R}Y_{L_H M_H}(\theta_H,\phi_H)Y_{L_R M_R}(\theta_R,\phi_R)
    \label{basis}
\end{eqnarray}
represent a quite suitable basis for the diagonalization of the Hamiltonian (\ref{hamiltonian}).
The action of the space reflection operator $\hat P$ on these functions
provides the corresponding parity quantum number
\begin{eqnarray}
    \hat P | L_H L_R; I M\rangle=(-1)^{L_H+L_R}| L_H L_R; I M\rangle \,.
\end{eqnarray}
Evidently, the molecule states are characterized by positive and negative parity numbers.
In our consideration, however, the first fragment is assumed to have  the axially-symmetric density
distribution (shape). Therefore, the wave function of
this fragment owns the inversion symmetry, which yields
\beq
\hat P Y_{L_H M_H}(\theta_H,\phi_H)=Y_{L_H M_H}(\pi-\theta_H,\phi_H+\pi)=(-1)^{L_H}Y_{L_H M_H}(\theta_H,\phi_H)\,.
\eeq
The spectrum of this fragment contains states only with even
or only odd values of the orbital momentum
with positive or negative parity values, respectively.
In our analysis we consider excited  states of the
axially-symmetric fragment, characterized  by a positive
parity quantum number with $L_H=0,2,4,\ldots$.
On the other hand, the total sum $L_H+L_R$ will determine
the parity of the molecule state, where $L_R=0,1,2,3, \ldots$.

The Hamiltonian $\hat H_{rot}$ is diagonal in the basis (\ref{basis})
\begin{eqnarray}
\label{anexp}
    \hat H_{rot}| L_H L_R; I M\rangle=\left (\frac{\hbar^2 L_H(L_H+1)}{2 \Im_H}+\frac{\hbar^2 L_R(L_R+1)}{2 \Im_R} \right )| L_H L_R; I M\rangle \,,
\end{eqnarray}
while the matrix elements of the interaction $V_{int}$ are expressed in the following analytical form
\begin{eqnarray}
&&\langle |f_1 f_2;I_f M_f|V_{int}| i_1 i_2; I_i M_i\rangle=\delta_{I_i I_f} \delta_{M_i M_f}\frac{C(\beta_2)}{3} \nonumber \\
&&\times \left (\delta_{i_1 f_1}\delta_{i_2 f_2} -
(-1)^{f_1+i_2+I_f}\sqrt{(2i_1+1)(2i_2+1)}C^{f_1 0}_{i_1 0 2 0}C^{f_2 0}_{i_2 0 2 0} \begin{Bmatrix} i_2 &f_2 &2\\
f_1 & i_1 & I_f\end{Bmatrix}\right ).
    \label{me}
\end{eqnarray}

The wave function of $n$-th excited state of the molecule
with the orbital momentum $L$ and its projection $M$ on the laboratory $z$-axis
can be presented as a superposition of the bipolar spherical functions (\ref{basis}):
\begin{eqnarray}
    \Psi_{n I M}=\sum_{L_H L_R}a^{(n I)}_{L_H L_R} \left [Y_{L_H}(\theta_H,\phi_H) \times Y_{L_R}(\theta_R,\phi_R)\right ]_{(IM)}.
    \label{bipol}
\end{eqnarray}
We recall that the ${\tilde z}$-axis  of the molecule (the body--fixed) coordinate system coincides with the symmetry axis of the deformed fragment
(see Fig.\ref{fig2}). Additionally, the vector ${\bf R}$ (that connects the fragment centers)  lies in the $O'\tilde{x}'\tilde{z}'$ frame of the body--fixed system.
Consequently, we rewrite the expression (\ref{basis})
with the aid of the Euler angles $\tilde{\Omega}=\{\phi_R,\theta_H,\phi_H \}$ (which describe the orientation of the body--fixed coordinate system), and the angle $\epsilon$, describing the orientation of ${\bf R}$ with respect to the body--fixed coordinate system.
We have the following relations between the angles $\Omega_H$ and $\Omega_R$ on the one side and the angles $\tilde \Omega$ and $\epsilon$ from the other side:
\begin{eqnarray}
\label{Kproj1}
    Y_{L_H M_H}(\theta_H,\phi_H)&=&
\sqrt{\frac{2L_H+1}{4\pi}}D^{L_H *}_{M_H 0}(\tilde \Omega)\,, \\
     Y_{L_R M_R}(\theta_R,\phi_R)&=&\sum_{k_1}D^{L_R *}_{M_R k_1} (\tilde \Omega)Y_{L_R k_1}(\epsilon,0)\,,
     \label{Kproj2}
\end{eqnarray}
which yield the expression for the bipolar function
\begin{eqnarray}
\label{expansion}
\left [Y_{L_H}(\theta_H,\phi_H) \times Y_{L_R}(\theta_R,\phi_R)\right ]_{(IM)}& = &
  \sqrt{\frac{2L_H+1}{4\pi}}  \sum_K C^{IK}_{ L_H 0 L_R K}D^{I *}_{M K} (\tilde \Omega)Y_{L_R K}(\epsilon,0)\,
\end{eqnarray}
as a superposition over states with different projection $K$ of the orbital momentum on the
body--fixed $O\tilde z$-axis. Evidently, this result implies the nonaxial shape of the molecule, that is requires numerical diagonalization of the molecular Hamiltonian. In some physically important cases, however, simple approximate solutions can be constructed. We outline briefly  two such cases.

\begin{enumerate}
\item  $L_R \approx I$.
The deformation of the first fragment is particularly small: $\beta_2 \sim 0$ and,
therefore, we have $\Im_H \approx 0$.
Since the lowest  rotational states of the fragment are located high in energy,
the rotational states of the fragment can be neglected in the description of the yrast spectrum of the molecule.  Additionally, it follows that in Eq.(\ref{hamiltonian})
the contribution of $V_{int}$ is negligible as  well due to $ C(\beta_2, R_m) \approx 0 $.
From these two conditions it follows that  the yrast spectrum of the molecule is characterized by the  moment of inertia in the rolling limit
\begin{equation}
\Im \approx \Im_R \gg \Im_H \approx 0\,.
\end{equation}
Consequently, the spectrum of a nuclear molecule is defined by
the analytical expression (\ref{anexp}), which provides the rotational energy determined by the term $$\frac{\hbar^2 L_R(L_R+1)}{2 \Im_R}$$ alone.  As evident from Eq.\eqref{expansion}, $L_H \approx 0$ leads to $K \approx 0$. The molecule in this case has practically an axially-symmetric shape.

\item $L_R \approx 0$.
The deformed fragment is characterized by a significant deformation $\beta_2 \gg 0$ which yields the following relations: $\Im_H \gg \Im_R$, and  the potential stiffness
$C(\beta_2,R_m) \gg 1$. In this case the tunneling  of the spherical fragment
between two pole-to-pole configurations in the coordinate $\epsilon$ is
almost forbidden due to the large barrier at $\epsilon=\pi/2$.
It is blocked and can only oscillate with a small amplitude around the pole-to-pole
configuration. Consequently, the excitation spectrum of the molecule consists of the rotational and vibrational spectra. The rotational spectrum is determined
by the rotation of the molecule as a whole with the moment of inertia
\beq
\label{tmoi}
\Im=\Im_H+\Im_R.
\eeq
The vibrational excitations are determined by the oscillation of the spherical fragment.

From the above discussion  and the structure of the wave function given by Eqs.(\ref{basis}), (\ref{expansion}),
we conclude that for the yrast-state $L_R \approx 0$ and $K \approx 0$. Moreover, taking into account
the asymptotic value of $Y_{L \pm K}(\epsilon) \sim (\epsilon/2)^K$,
it follows that only small values of $K$ can contribute to the structure of the low-lying states.
The $K$ values are generated by the vibrational motion of the molecule in
the coordinate $\epsilon$ around
pole-to-pole configuration.
Consequently, we assume that the low-energy spectrum of the molecule will constitute of several bands
characterized by the moment of inertia close to the sticking moment of inertia $\Im \approx \Im_H+\Im_R$.
\end{enumerate}
In this limit the nuclear molecule has nearly axially-symmetric shape. Below we consider the latter case
in more detail in order to
obtain analytical expressions, that
will elucidate the structure of the  low-energy spectrum of the molecule.

\section{Analytical solutions}
\label{ansols}

\subsection{The Hamiltonian in the body--fixed frame}
\label{3a}

Within the assumption of a weak overlap between the fragments, we have obtained the kinetic energy of our system in the laboratory frame (see Sect.\ref{Sec_ke}).
In the body-fixed frame the classical expression for the kinetic energy is derived in Appendix \ref{qkinen}.
Applying the Pauli quantization procedure \cite{Pauli} (see Eq.(\ref{ap4})), we obtain
the kinetic energy operator in the  form:
\bea
&&\hat T = \frac{\hbar^2}{2 \Im_{H}} \left [\hat I^2-\hat I_{3}^2+ 2 \cot{\epsilon} \hat I_1\hat I_3 +2 i \hat I_2 \frac{\partial}{\partial \epsilon}\right]\nonumber \\
\label{k2}
&-& \frac{\hbar^2}{2  \Im_b}\left [ \frac{\partial^2}{\partial \epsilon^2}+\cot{\epsilon} \frac{\partial}{\partial \epsilon} - \frac{ \hat I_3^2 }{\sin^2\epsilon}\right ]\,,
\eea
where
$\Im_{R}=\mu R^2_{m}\,,\Im_b=\Im_{H}\Im_{R}/(\Im_{H}+\Im_{R})\,.$
In Eq.~\eqref{k2}, the intrinsic components of the total angular momentum
are:
\begin{eqnarray}
&&\hat I_1= -i \left \{ - \frac{\cos{\theta_3}}{\sin{\theta_2}}\frac{\partial}{\partial \theta_1}+\sin{\theta_3}\frac{\partial}{\partial \theta_2}+\cot{\theta_2}\cos{\theta_3}\frac{\partial}{\partial \theta_3}\right \}\,, \nonumber \\
&&\hat I_2= -i \left \{ - \frac{\sin{\theta_3}}{\sin{\theta_2}}\frac{\partial}{\partial \theta_1}+\cos{\theta_3}\frac{\partial}{\partial \theta_2}-\cot{\theta_2}\sin{\theta_3}\frac{\partial}{\partial \theta_3}\right \}\,, \nonumber \\
&&\hat I_3= -i \frac{\partial}{\partial \theta_3}\,,
\end{eqnarray}
and
\bea
\hat I^2&=&\left \{ -\frac{\partial^2}{\partial \theta_2^2}-\cot{\theta_2}\frac{\partial}{\partial \theta_2}-\frac{1}{\sin{\theta_2}^2}\left ( \frac{\partial^2}{\partial \theta_1^2}+\frac{\partial^2}{\partial \theta_3^2}\right )\right. \nonumber \\
&+&\left.2\frac{\cos{\theta_2}}{\sin{\theta_2}^2}\frac{\partial^2}{\partial \theta_1 \partial \theta_3} \right \}\,.
\eea

The kinetic energy \eqref{k2} yields together with the potential term $C/2 \sin^2{\epsilon}$ (see \eqref{potential})   the total Hamiltonian of the dinuclear system:
\begin{eqnarray}
\label{ap66a}
&&\hat H= {\hat H}_{rot}+{\hat H}_{bend}+ V_{int}= {\hat H}_{0}+V_{int}\,,  \\
\label{hr1}
&&{\hat H}_{rot}=\frac{\hbar^2}{2  \Im_H}(\hat I^2-2 \hat I^2_3)\,,\\
\label{hr2}
&&{\hat H}_{bend}=-\frac{\hbar^2}{2 \Im_b}
\left [\frac{1}{\sin{\epsilon}}
\frac{\partial}{\partial\epsilon}\sin{\epsilon}\frac{\partial}{\partial\epsilon}
 \right ] +
\frac{\hbar^2}{2 \Im_b}\frac{\hat I^2_3}{\sin^2{\epsilon}}
+
\frac{1}{2}C(\beta_2,R_m) \sin^2{\epsilon} \,,\\
\label{hr3}
&&V_{int}=\frac{\hbar^2}{2 \Im_H} \left [
 \cot{\epsilon}  \left ( \hat I_1 \hat I_3+\hat I_3 \hat I_1\right)+2 i \hat I_2
\frac{1}{\sqrt {\sin{\epsilon}}} \frac{\partial}{\partial \epsilon}\sqrt
{\sin{\epsilon}}\right ].
\end{eqnarray}

We recall that the potential energy of the molecule is determined by a sum of the nuclear $U_N$ and the Coulomb $U_C$ energies, defined in the
touching configuration at distance $R=R_m(\epsilon)$ (see Sec.\ref{poten}):
\beq
U(\epsilon)\equiv U(\beta_2,R_m(\epsilon),\epsilon)=
U_N(\beta_2,R_m(\epsilon))+U_C(\beta_2,R_m(\epsilon),\epsilon)\,.
\eeq
The nuclear potential is proportional to the volume of the overlapping region of fragment densities in the touching configuration, that is, it is not strongly dependent on the angle $\epsilon$ \cite{Blocki1977,AR2024}.
The nuclear potential acts as a "nuclear glue" that holds two fragments together. Consequently, the stiffness parameter $C$ can be estimated from the difference between the the potential energies
of the pole-to-pole and the equatorial configurations (see discussion before Eq.\eqref{potential}), determined by the Coulomb interaction alone. Taking into account the results presented in Appendix \ref{calc3},  we obtain
\begin{eqnarray}
\label{stiff_approx}
    C=2 [U_C(\epsilon=\pi/2)-U_C(\epsilon=0)] \approx 3 \sqrt{\frac{5}{4\pi}}U_{C}(\beta_2=0)
    \frac{R_{1}}{R_{1}+R_{2}} \left ( 1-\frac{3}{5} \frac{R_{1}}{R_{1}+R_{2}} \right)\beta_2\,.
\end{eqnarray}
Here, this result is obtained, keeping only linear term in the deformation parameter of the deformed fragment $\beta$.

\subsection{Basis states and symmetries}
\label{symm}

The Hamiltonian $\hat H_0$ posses rotational and vibrational degrees of freedom. Assuming that
its wave function has the following form
\bea
\label{WF_H0}
\Psi^\pi_{IMKn}=|IMK\rangle \times \phi_{nK}(\epsilon)\,,\\
|IMK\rangle =\sqrt{\frac{2I+1}{16\pi}}D^{I*}_{MK}(\theta_1,\theta_2\,.\theta_3)\,,
\eea
we obtain the equation for the intrinsic wave function
\begin{eqnarray}
\hat H_{bend}\phi_{nK}(\epsilon)=E_{nK}\phi_{nK}(\epsilon)\,.
    \label{basis_vib}
\end{eqnarray}
Here,  $I$, $M$, and $K$ are the quantum numbers of the total angular momentum and its projections on the laboratory axis $z$ and the intrinsic axis $\tilde z$, respectively. The parity quantum number is denoted as $\pi$.
It should be pointed out that
for the considered system the parity operation $\hat \pi$ is equivalent to the reflection $\hat P$ with respect to the intrinsic plane $\tilde x \tilde y$ and the rotation $\hat R_3 =\exp{(i \pi \hat I_3)}$  around the axis $\tilde z$ by the angle  $\pi$. Consequently, we have
    \begin{eqnarray}
     \label{parity}
    \hat \pi &=&\hat P \hat R_3\,,\\
    \label{exc}
    \hat \pi  \Psi^\pi_{IMKn}(\theta_1, \theta_2, \theta_3,\epsilon)&=&p\Psi^\pi_{IMKn}(\theta_1, \theta_2, \theta_3-\pi,\pi-\epsilon)\,.
\end{eqnarray}

Since the Hamiltonian $\hat H_{bend}$ is symmetric with respect to the reflection $\hat P$, the eigenvalues
$\phi_{nK}(\epsilon)$ are either symmetric or antisymmetric functions with respect to the transformation $\epsilon \rightarrow \pi-\epsilon$. The symmetric and antisymmetric states interleave each other, with symmetric states being lower in energy.
Enumerating the states starting from $n=0$, we obtain
\begin{eqnarray}
\hat P \phi_{nK}(\epsilon)=(-1)^n \phi_{nK}(\epsilon).
    \label{reflection}
\end{eqnarray}
Using the property
\bea
\hat R_3 D^I_{M,K}(\theta_1,\theta_2,\theta_3)=(-1)^K D^I_{M,K}(\theta_1,\theta_2,\theta_3),
    \label{R3}
\eea
we have
\begin{eqnarray}
\hat \pi |IMK\rangle \times \phi_{nK}(\epsilon)=(-1)^{n+K}|IMK\rangle \times \phi_{nK}(\epsilon)
    \label{evpar}.
\end{eqnarray}
As a result, we found that eigenstates (\ref{WF_H0}) have a good parity quantum number defined by the relation
$p=(-1)^{n+K}$.

Our eigenstates are obtained for the particular choice of the intrinsic system. However, the orientation of the axes of the internal system is determined arbitrarily.
As a matter of fact, to have an uniquely defined  wave function in the laboratory coordinate system it should be invariant with respect to the renaming of axes of the intrinsic system at a particular rotation: $\Psi (\hat{\mathcal{L}}_i \alpha)=\Psi (\alpha)$, where $\alpha=(\theta_1,\theta_2,\theta_3,\epsilon)$.
To realize this invariance let us consider two operators
\begin{eqnarray}
\label{fs}
   \mathcal{L}_1&=& \hat \pi \hat R_2=\hat \pi \exp{(i \pi \hat I_2)}\,,\\
\label{ss}
   \mathcal{L}_2&=&\hat P \hat R_1= \hat P \exp{(i \pi \hat I_1)}.
\end{eqnarray}
The transformation (\ref{fs}) contains the product of the parity operation
$\hat \pi$ and rotation of the system around the intrinsic axis $\tilde y$ on angle $\pi$.
Evidently, from this transformation it follows that the
fully symmetric state with a definite parity should have the following form
\beq
\Psi^S_{IMK\pi n}= \frac{1}{\sqrt{2}}\left (|IMK\rangle+  \mathcal{L}_1 |IMK\rangle \right )\times \phi_{nK}(\epsilon),
\eeq
Taking into account Eq.(\ref{parity}) and the relation
\beq
   \hat R_2 D^I_{M,K}(\theta_1,\theta_2,\theta_3)=(-1)^{I+K} D^I_{M,-K}(\theta_1,\theta_2,\theta_3)\,,
\eeq
we obtain the  symmetrized eigenstates of $\hat H_0$ in the following form
\begin{eqnarray}
 \Psi^S_{IMK\pi n}= \frac{1}{\sqrt{2}}\left (|IMK\rangle+p(-1)^{I} |IM-K\rangle \right )\times \phi_{nK}(\epsilon),
 \label{WF_sym}
\end{eqnarray}
where $p=(-1)^n$. Similarly, the transformation  (\ref{ss})
should define the symmetric state with a definite parity, i.e.,
\beq
\Psi^S_{IMK\pi n} = \frac{1}{\sqrt{2}}\left (|IMK\rangle+  \mathcal{L}_2 |IMK\rangle \right ) \times \phi_{nK}(\epsilon)\,.
\eeq
Taking into account Eq.(\ref{reflection}) and the following property
\beq
 \hat R_1 D^I_{M,K}(\theta_1,\theta_2,\theta_3) =  (-1)^{I} D^I_{M,-K}(\theta_1,\theta_2,\theta_3)\,,
\eeq
we obtain the eigenfunction (\ref{WF_sym}) which is invariant also under the operation $\mathcal{L}_2$.
As a result, the eigenfunction Eq.(\ref{WF_sym})  has the
following good quantum numbers:
the angular momentum $I$, its projections on the laboratory z-axis $M$ and the intrinsic 3rd axis $K$,
the parity $p=(-1)^{n+K}$.

Below we consider two cases, where the eigenstates of
the Hamiltonian $\hat H_{bend}$ can be found analytically and the effect of the interaction $V_{int}$ can be taken into account by means of the perturbation theory.

\subsection{Roto-vibrations around the pole-to-pole configuration}
\label{an1}

We recall that the potential energy of the molecule has
a form of a double well with  two minima at $\epsilon=0$ and $\epsilon=\pi$, separated by the barrier at $\epsilon=\pi/2$.
Neglecting the tunneling through the barrier, we study the vibrations around $\epsilon=0$
 and $\epsilon=\pi$ independently of each other. Since the barrier is quite high,  there is a degeneracy of two excited levels. A superposition of these two levels yields good parity states that will be used for
the eigenvalue problem of the considered configuration.

Assuming $\sin \epsilon \approx \epsilon$ and $\sin \epsilon \approx \pi- \epsilon$, we expand
 the Hamiltonian \eqref{ap66a} with respect to $\tilde \epsilon=\epsilon$ or  $\tilde \epsilon=\pi-\epsilon$ and obtain:
\begin{eqnarray}
\label{ap66}
&&H= H_{rot}+H_{bend}+ V_{int}\,,  \\
&&H_{bend}=\frac{\hbar^2}{2 \Im_b}
\left [-\frac{1}{\tilde \epsilon}\frac{\partial}{\partial\tilde \epsilon}\tilde \epsilon\frac{\partial}{\partial\tilde \epsilon}
+\frac{ \hat I^2_3}{\tilde \epsilon^2}\right ]
+\frac{1}{2}\left[C(\beta_2,R_m)+\frac{\hbar^2}{15 \Im_b}\hat I^2_3\right]\tilde \epsilon^2 \,, \nonumber \\
&&V_{int}=\frac{\hbar^2}{2 \Im_H} \left [
\frac{1}{\tilde \epsilon}  \left (\hat I_1\hat I_3+\hat I_3 \hat I_1\right)+2 i \hat I_2
\frac{1}{\sqrt {\tilde \epsilon}} \frac{\partial}{\partial \tilde \epsilon}\sqrt
{\tilde \epsilon}\right ]+\frac{\hbar^2}{6\Im_b}\tilde \epsilon\frac{\partial}{\partial \tilde \epsilon}\,,
\label{app66c}
\end{eqnarray}
where the Hamiltonian $H_{rot}$ is defined by Eq.(\ref{hr1}).
Hereafter, for the sake of convenience we use $C\equiv C(\beta_2,R_m)+\hbar^2 K^2/15 \Im_b\approx C(\beta_2,R_m)$. This approximation is justified quite well, since $\hbar^2 K^2/15 \Im_b\ll 1$ ($K$ is small).
In this approximation the spectrum of the Hamiltonian $\hat H_{bend}$
is simply the spectrum of the harmonic oscillator. As a result,
the eigenstates of the Hamiltonian $\hat H_{bend}$ can be found analytically
\begin{eqnarray}
&&\hat H_{bend}\phi_{nK}(\tilde \epsilon)=\hbar \omega (2 n+|K|+1)\phi_{nK}(\epsilon)  \,, \nonumber \\
&&\phi_{nK}(\tilde \epsilon)=\sqrt{\frac{2 n!}{(K+n)!}} \frac{\tilde \epsilon^K }{\epsilon_0^{K+1}}
\exp \left (-\frac{\tilde \epsilon^2}{2\epsilon_0^2} \right )L_n^{(|K|)}\left (\frac{\tilde \epsilon^2}{\epsilon_0^2} \right )\,,
    \label{laguerre}
\end{eqnarray}
where $L_n^{(|K|)}$ are the generalized Laguerre polynomials.
Here the quantity
\beq
\label{e0}
\epsilon_0^2=\hbar/\sqrt{C \Im_b}
\eeq
is the standard deviation of the system  from the axial symmetry,
and the frequency of the vibrations in $\epsilon$ degree of freedom is defined as
\begin{equation}
\label{om}
\omega=\sqrt{C/\Im_b}\equiv \hbar/(\Im_b\epsilon^2_0)\,.
\end{equation}
The symmetric or antisymmetric eigenstates of $\hat H_{bend}$ can be constructed from two degenerate solutions corresponding to the vibrations around $\epsilon \approx 0$ and $\epsilon \approx \pi$ as
\begin{eqnarray}
\Phi_{nKp}(\epsilon) = \frac{1}{\sqrt{2}}\left (\phi_{nK}(\tilde \epsilon=\epsilon)+p \psi_{nK}(\tilde \epsilon=\pi-\epsilon ) \right ),
    \label{sym_Laguerre}
\end{eqnarray}
where $p=\pm 1$ corresponds for the solutions which are symmetric (+) or asymmetric (-) with respect to reflection operation $\hat P$. As a result, we construct the eigenstates  of
the Hamiltonian $\hat H_0=H_{rot}+H_{bend}$ in the form
\begin{eqnarray}
    \Psi^{\pi}_{IMKn}= \frac{1}{\sqrt{2}}\left (
    |IMK\rangle +p(-1)^I|IM-K\rangle\right )\Phi_{nKp}(\epsilon)\,.
\end{eqnarray}
These  states are characterized by a good parity and preserve the symmetry with respect to $\mathcal L_1$ and $\mathcal L_2$ operations. Consequently, we have the following eigenvalues:
\begin{eqnarray}
\label{zero}
 E_{I,M,K,n}^{(0)}&=&\frac{\hbar^2}{2 \Im_H}\left[I(I+1)-2
K^2 \right] +\hbar \omega  (2 n+|K|+1 )\,.
\end{eqnarray}
This expression resembles the results of Refs.\cite{Uegaki1993,Uegaki2012}, that are obtained without  the coupling $V_{int}$ between rotational and vibrational motion   (see discussion in Sec.\ref{intro}). Our aim is to go beyond this approximation and to analyze the effect of the coupling Eq.(\ref{app66c}) for actinide and transactinide mass regions.
With the aid of Eqs.\eqref{stiff_approx},\eqref{invmb}
it follows from Eq.\eqref{e0} that $\epsilon_0 \ll 1$. Therefore, $\epsilon_0$ can be used as a perturbative parameter in the expansions of various physical quantities of the nuclear molecule in the pole-to-pole limit. In particular, we  consider the contribution of the
interaction Eq.(\ref{app66c}) up to the second order in the perturbation approach (see Appendix \ref{calc2}) to the total energy.
This contribution mixes the basis states with different $K$ and $n$ quantum numbers.
This mixing adds small components with $\Delta K \approx K \pm 1$ and $\Delta n \approx n \pm 1$, distributed around $K$ and $n$. Therefore, we use the asymptotic quantum numbers $K$ and $n$ to label the quantum states.
With the aid of Eq.(\ref{4order}) and the  wave function
\begin{eqnarray}
\Psi^{(n,K)}_{I,M}=\sum_{k=K \pm 1}\sum_{n = n \pm 1}
a(n,k)|I M k  \pi\rangle \Phi_{nKp}(\epsilon)
\label{analyt_wf}\,,
\end{eqnarray}
we obtain for the total energy of the molecular state the following result
\bea
E^{(n,K)}_{I,M}\approx\frac{\hbar^2}{2 \Im}[I(I+1)-2K^2]
+\frac{\hbar^2}{\Im_b \epsilon_0^2}  (2 n+|K|+1 )+\frac{\hbar^2}{6 \Im_b}\,,
\label{ap13}
\eea
expressed by means of the asymptotic quantum numbers $n$ and $K$.
Since the  spherical fragment is situated at the pole of the heavy fragment, the
total moment of inertia $\Im$ is expected in the sticking limit. In contrast to Eq.(\ref{zero}), the rotational term in Eq.(\ref{ap13}) is determined by the total moment of inertia of the molecular system.

\begin{figure}[t]
\includegraphics[scale=0.99]{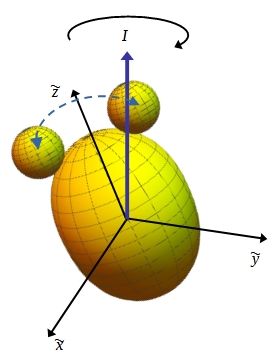}
\vspace{-.5cm} \caption{Sketch of the pole-to-pole vibrations.}
\label{num3}
\end{figure}
 The spherical fragment (nucleus) oscillates around the symmetry axis of the heavy fragment (nucleus).
This oscillation can be presented as a superposition of two independent vibrations in
the planes perpendicular to the plane $O{\tilde x}{\tilde y}$. Each vibration contributes $\hbar \omega(n+1/2)$ ($\hbar \omega = \hbar^2/(\Im_b\epsilon_0^2$)) to the energy of the pole-to-pole configuration (see Eq.(\ref{ap13})).
Indeed, using the relation between the second quantization
representation and the polar coordinate representation, we
have for two oscillators
\beq
E_{n_{+},n{-}}=\hbar\omega_0 (n_{+}+1/2) + \hbar \omega_0 (n_{-}+1/2)=\hbar \omega_0(2n+|K|+1)
\eeq
with $n_++n_-=2n+|K|$ (see, e.g., Appendix in \cite{bir}).
Additionally, these vibrations lead to the
nonaxial shapes and generate the nonzero $K$-projection of the total angular momentum.

\subsection{Roto-vibrations  in the equatorial region}
\label{an2}

\begin{figure}[t]
\includegraphics[bb=0 0 380 342,scale=0.55]{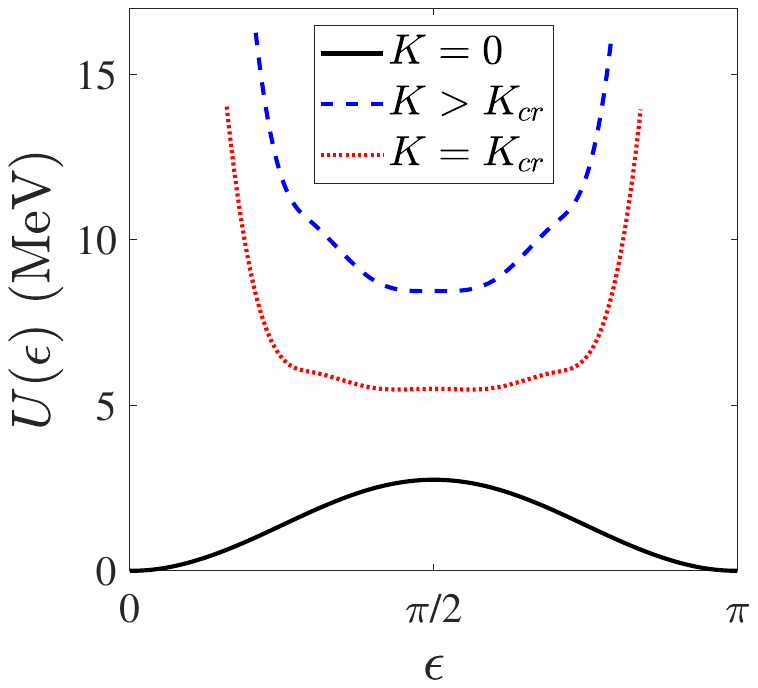}
\vspace{-.5cm} \caption{Sketch of the effective potential $V(\epsilon) \equiv U(\epsilon)$  as a function of the angle $\epsilon$ (see Eq.~(\ref{eff_pot})) for different values of $K$: K=0 (black solid line); $K=K_{cr}$ (red dotted line), and $K>K_{cr}$ (blue dashed line). For the illustration, the following parameters $C=5.7$ MeV and $\Im_b=7.1$ $\hbar^2/$MeV have been used (see details in Sec.~\ref{cnum}).}
\label{efpot}
\end{figure}

Another  curious analytical solution for the Hamiltonian (\ref{k2}) can be obtained if  $K$ value is larger than some critical value $K_{cr}$.
In this case the properties of the molecule will be defined by
the effective potential energy (see Eq.(\ref{hr2})), which includes the
competition between the harmonic trap and the centrifugal-like term:
\begin{eqnarray}
U(\epsilon) = \frac{C}{2} \sin^2{\epsilon}+\frac{\hbar^2}{2 \Im_b \sin^2{\epsilon}}K^2.
\label{eff_pot}
\end{eqnarray}
Note that for $K=0$, potential energy has  two minima at $\epsilon=0$ and $\pi$ separated by the barrier at $\epsilon=\pi/2$.
With the $K$ increase the potential energy owns the other shape with a single minimum at $\epsilon=\pi/2$ (see Fig.~\ref{efpot}).
The extremum conditions for the potential $U(\epsilon)$ determine the critical value:
\beq
|K_{cr}| = \frac{\sqrt{C \Im_b}}{\hbar} =1/\epsilon^2_0\,.
    \label{Kcr}
\eeq
In other words,
with the increase of $K$ the two minima structure of the potential energy transforms
to the one minimum at $K_{cr}$, i.e., a type of the "phase" transition takes place. At this point quantum fluctuations of the pole-pole configurations tends to zero, and, consequently, there is a freedom for the light fragment to move over the surface of the heavy fragment. With
$K\geq K_{cr}$ there is  the onset of the rotation of a small fragment around the heavy fragment.
This motion binds the small fragment in the equator configuration (\ref{eff_pot}).
For $K \gg K_{cr}$ we expand the Hamiltonian (\ref{ap66a})
in $\tilde \epsilon=\pi/2-\epsilon$ and obtain
\begin{eqnarray}
&&H= H_{rot}+H_{bend}+ V_{int}=H_0+ V_{int}\,,
\label{app66d}\\
&&H_{bend}=-\frac{\hbar^2}{2 \Im_b}
\frac{\partial^2}{\partial\tilde \epsilon^2}
+\frac{1}{2}\left (\frac{\hbar^2 K^2}{\Im_b}-C \right ) \tilde \epsilon^2+\frac{1}{2}\left (\frac{\hbar^2 K^2}{\Im_b}+C \right )\,, \nonumber \\
&&V_{int}=-\frac{\hbar^2}{2 \Im_H} \left [
\tilde\epsilon \left ( \hat I_1 \hat I_3+\hat I_3 \hat I_1\right)+2 i \hat I_2
\frac{\partial}{\partial \tilde\epsilon} \right ] .
    \label{oscil_solution}
\end{eqnarray}
Here the Hamiltonian $H_{rot}$ is defined by Eq.(\ref{hr1}).
The  Hamiltonian $H_0$ in Eq. (\ref{app66d})
can be diagonalized analytically with the aid of the wave functions:
\begin{eqnarray}
\label{ap7a}
&&\Phi_{IMKn}= |IMK\rangle \times \phi_{n}(\tilde\epsilon)\,,\\
&&|IMK\rangle=  D^{I*}_{M,K}(\theta_1,\theta_2,\theta_3)\,,\nonumber \\
&&\phi_{n}(\tilde\epsilon)=\frac{1}{\pi^{1/4} \tilde \epsilon_0^{1/2}} \frac{1}{\sqrt{2^n n!}}
\exp \left (-\frac{\tilde\epsilon^2}{2\tilde \epsilon_0^2} \right )H_n\left (\frac{\tilde \epsilon}{\tilde\epsilon_0} \right )\,, \nonumber
\end{eqnarray}
where $K$ is a good quantum number.
Here $H_n$ are the Hermite polynomials,
and the standard deviation of the system
from $\epsilon=\pi/2$ is defined by the variable $\tilde \epsilon_0^2=\hbar/\sqrt{\left (\frac{\hbar^2 K^2}{\Im_b}-C \right ) \Im_b}=1/\sqrt{K^2-K^2_{cr}}$.
We recall that the action of the parity operator (\ref{parity}) on the state \eqref{ap7a} is equivalent to the change of $\epsilon$ to $\pi-\epsilon$ and to the rotation $R_3=\exp{i \pi \hat I_3}$ around the intrinsic axis $z'$ on angle $\pi$ (see Eq.{\ref{exc})).  From the properties of Hermite polynomials it follows that
\bea
\label{Hermitean}
\tilde \epsilon =\pi/2-\epsilon \rightarrow \epsilon-\pi/2 = -\tilde \epsilon  \nonumber \\
H_n(-\tilde\epsilon)= (-1)^n H_n(\tilde \epsilon)\,.
\eea
Taking into account Eq.(\ref{R3}), we obtain the parity quantum number for the state $\Phi_{IMKn}$ as  $\pi=(-1)^{n+K}$.
Consequently, for each quantum number $K$ the model provides sequences of bands with alternating parities. For example, the positive parity band is characterized by even values of the sum $K+n$, while the negative one with odd values of this sum with the same $K$.

The states (symmetrized with respect to the operations $\mathcal{L}_1$ and $\mathcal{L}_2$) have the form
\bea
\label{basis_sym}
&&\Phi^{S}_{IMKn}= |IMK\pi\rangle \times \phi_{n}(\tilde\epsilon)\,,\\
&&|IMK\pi\rangle= \frac{1}{\sqrt{2(1+\delta_{K0})}}\left \{ |IMK\rangle+ (-1)^{n+I}|IM-K\rangle \right \}\,,\nonumber \\
&&\phi_{n}(\tilde\epsilon)=\frac{1}{\pi^{1/4} \tilde \epsilon_0^{1/2}} \frac{1}{\sqrt{2^n n!}}
\exp \left (-\frac{\tilde\epsilon^2}{2\tilde \epsilon_0^2} \right )H_n\left (\frac{\tilde \epsilon}{\tilde\epsilon_0} \right )\,, \nonumber
\eea

The eigenvalues of the Hamiltonian $H_0$ are:
\begin{eqnarray}
\label{zero1}
 E_{I,M,K,n}^{(0)}&=&\frac{\hbar^2}{2 \Im_H}\left[I(I+1)-2
K^2\right] +\hbar {\tilde\omega}  (n+1/2 )+
\frac{1}{2}\left(\frac{\hbar^2 K^2}{\Im_b}+C \right ) ,
\end{eqnarray}
where ${\tilde{\omega}}=
\sqrt{\left (\frac{\hbar^2 K^2}{\Im_b}-C \right ) /\Im_b}=\omega\sqrt{K^2/K^2_{cr}-1}$ is the frequency of the $\epsilon$-vibrations.

For $K \gg K_{cr}$ we have $\hbar \tilde \omega = \hbar^2 K/\Im_b$, and,
taking into account the definition (\ref{invmb}),
the eigenvalues \eqref{zero1} transforms to the following forms
\begin{equation}
E^{(0)}_{I,M,K,n}=\frac{\hbar^2}{2 \Im_H}\left [I(I+1)-K^2 \right ]+\frac{\hbar^2 K^2}{2 \Im_R} +
\frac{\hbar^2 K (n+\frac{1}{2}) }{\Im_b}+\frac{C}{2}.
\label{approx_question}
\end{equation}

In this approximation the system manifests a tilted rotation, characterized by  the rotation around axes $x$ and $y$ with moments of inertia $\Im_1=\Im_2=\Im_H$ and the rotation around the axis $z$ with the moment of inertia $\Im_3=\Im_R$.
The third term corresponds to the harmonic oscillations, i.e., it describes the precession of the projection of the angular momentum $K$ around  the direction of the
total angular momentum $I$ with the frequency $\tilde \omega$. This precession is determined by
the reduced moment of inertia $\Im_b=(\Im_R \times \Im_H)/(\Im_R+\Im_H)$.

Let us estimate the modifications to the energies $E_{I,M,K,n=0}$ due to the coupling term $V_{int}$, evaluated in the perturbation approach.
The matrix elements of the coupling $V_{int}$ with respect to the rotational part are given by the expressions \eqref{ap9},\eqref{ap10}, while those for the vibrational part are given by
the expressions \eqref{vb1} and \eqref{vb2}. Considering $|K| \gg K_{cr}$,  $\epsilon_0 \approx 1/\sqrt{K}$, we obtain for the nonzero matrix elements
\begin{eqnarray}
\label{matel1}
&& \frac{2 \Im_H}{\hbar^2} \langle I M (K+1),n=1|\hat V_{int}|I M K,n=0\rangle= \nonumber \\
&&-\frac{\sqrt{L(L+1)-K (K+1)} \left[(2 K+1) \epsilon_0^2+2\right]}{2 \sqrt{2} \epsilon_0} \approx -\sqrt{2K(L(L+1)-K (K+1))}, \\
&&\frac{2 \Im_H}{\hbar^2}\langle I M (K-1),n=1|\hat V_{int}|I M K,n=0\rangle= \nonumber \\
&&-\frac{\sqrt{L(L+1)-K (K+1)} \left[(2 K-1) \epsilon_0^2-2\right]}{2 \sqrt{2} \epsilon_0} \approx 0.
    \label{matel}
\end{eqnarray}
In the same approximation, the energy denominator has the form
\begin{eqnarray}
\delta E& =&E^{(0)}_{I,M,K,n=0}-E^{(0)}_{I,M,K+1,n=1}\nonumber \\
&=&-\frac{\hbar^2}{\Im_H}-\frac{2 \hbar^2}{\Im_R}(K+1) =
 -\frac{2 \hbar^2 K}{\Im_R}-\frac{\hbar^2}{\Im_b \Im_R}(\Im_b+\Im_R)
\approx -\frac{2 \hbar^2 K}{\Im_R}\,, K\gg 1\,.
\label{denominator}
\end{eqnarray}
Using the expressions (\ref{matel1},\ref{denominator}), we obtain the second order correction to $E_{I,M,K,n=0}$ as
\begin{eqnarray*}
E_{cor} &=&\frac{|\langle I M (K+1),n=1|\hat V_{int}|I M K,n=0\rangle|^2}{\delta E} \nonumber \\
&=&-\frac{\hbar^2 \Im_R}{4 \Im_H^2} \left [ I(I+1)-K(K+1)\right ].
    \label{correction}
\end{eqnarray*}
Thus, for the total energy we  obtain
\begin{eqnarray}
\label{entot2}
E_{I,M,K,n}=\frac{\hbar^2}{2 \Im_H}\left [I(I+1)-K^2\right ] \left [1-\frac{\Im_R}{2\Im_H}\right ] +\frac{\hbar^2 K^2}{2 \Im_R}+
\frac{\hbar^2 K}{2 \Im_b}\left (n+\frac{1}{2} \right )+ \frac{C}{2}+\frac{ \hbar^2 K \Im_R}{4\Im_H^2}\,.
    \label{approx_question1}
\end{eqnarray}
Due to the condition $ \Im_R\ll \Im_H$ the last term in the energy (\ref{approx_question1}) can be neglected.

In order to delve deeper into the result (\ref{entot2}) we recall that a spherical fragment
can rotate  around the symmetry axis of the heavy fragment on the equatorial surface.
In this case the kinetic term has the following form
\beq
{\hat T} = \frac{{\hat I}_1^2}{2\Im_1} + \frac{{\hat I}_2^2}{2\Im_2}+\frac{{\hat I}_1^3}{2\Im_3} =
\frac{{\hat I}_1^2}{2(\Im_H+\mu R_x^2)} +
\frac{{\hat I}_2^2}{2(\Im_H+\mu R_y^2)}+\frac{{\hat I}_3^3}{2\Im_R}\,,
\eeq
where $R_{\tilde x}^2$ and $R_{\tilde y}^2$  are mean square distances between the center of light fragment and $\tilde x$ and $\tilde y$ axes, respectively.
Taking into account the symmetry with respect to the axes $\tilde x$ and $\tilde y$, during the course of this rotation, an average distances between the center of the spherical fragment  and axes $\tilde x$ and $\tilde y$ are equal, that is
$R_{\tilde x} = R_{\tilde y}$ and $R_{\tilde y}^2+R_{\tilde x}^2=R^2$. Here, it is taken into account that the spherical fragment is located on the equatorial plane.
Consequently, we have
\begin{eqnarray}
    \Im_{1(2)}=\mu R_{\tilde x(\tilde y)}^2+\Im_H =\frac{1}{2}\mu R^2+\Im_H =\frac{1}{2}\Im_R+\Im_H.
\end{eqnarray}
As a result,  the first term in Eq.(\ref{approx_question1}) can be understood as
a diagonal matrix element of the operator
\begin{eqnarray}
    \frac{I^2_1}{2 \Im_1}+\frac{I^2_2}{2 \Im_2}=\frac{I^2_1+I^2_2}{2 \Im_H+\Im_R} \approx \frac{\hat I^2-\hat I^2_3}{2\Im_H} \left [1-\frac{\Im_R}{2 \Im_H} \right ]\,,
\end{eqnarray}
in the basis wave functions (\ref{basis_sym}).


\begin{figure}[t]
\includegraphics[scale=2.5]{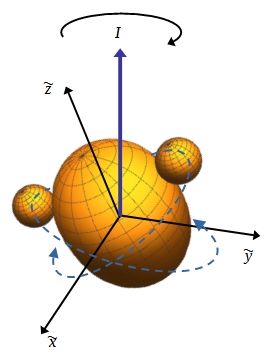}

\vspace{-.2cm} \caption{Sketch of the equatorial vibrations.}
\label{num33}
\end{figure}
Clustering of a spherical fragment (nucleus) and a heavy fragment (nucleus) near the equatorial surface results in the precessional motion of the heavy fragment symmetry axis around the angular momentum of the nuclear molecule  (see Fig.~\ref{num33}). For the sake of discussion the contact is assumed to occur in the upper hemisphere of the reaction. In the region of contact, the trajectory of the spherical fragment changes from the upper to the lower hemisphere, causing the angular momentum of the heavy fragment to move in the opposite direction. The motion of the spherical fragment  in the opposite direction results in the heavy fragment symmetry axis returning to its original direction, leading to precession of the nuclear molecule. The motion of the spherical fragment from the upper to the lower hemisphere and back is transformed into an effective oscillatory  motion (nutation) of the entire system relative to the direction of the angular momentum of the spherical fragment with a frequency
$\hbar\omega_{nut}=\hbar^2 K/(2\Im_b)$ (see  Eq.(\ref{approx_question1})).

Such a behavior resembles closely the Andronov-Hopf bifurcation \cite{andr,hopf}.
Indeed, at the first moment of the touching of a spherical fragment with the heavy one, the spherical fragment  can move along the equator of the ellipsoid with a constant speed without deviations. Upon reaching a certain critical speed, this motion along the equator loses stability.  The spherical fragment can no longer move in a straight line along the equator, but begins to oscillate (nutate) up and down relative to the equatorial plane, while continuing to move in a circle. In this case, the Andronov-Hopf bifurcation describes the loss of stability of the equatorial orbit and the birth of stable oscillatory orbits around it.

To illuminate the discussed phenomena we consider the semiclassical limit. In this limit the
ratio of the rotational frequency of the spherical fragment
\beq
\frac{\hbar^2K^2}{2\Im_R}\equiv \frac{\Im_R\omega_{rot}^2}{2}\Rightarrow \omega_{rot}=\frac{\hbar K}{\Im_R}
\eeq
and the nutation frequency $\omega_{nut}$ yields in the mass asymmetric molecule ($\alpha$-particle+$Pb$)  the following result
\beq
\label{rat}
\frac{\omega_{rot}}{\omega_{nut}}=\frac{2\Im_b}{\Im_R}=\frac{2\Im_H}{\Im_H+\Im_R}\approx 2\,,
\eeq
i.e., the rotation is twice faster than the nutation.
In the symmetric molecule (approximately, two equal mass fragments) the ratio gives
\beq
\frac{\omega_{rot}}{\omega_{nut}}\ll 1 \Leftarrow \Im_R\gg \Im_H\,.
\eeq
In other words, the nutation is much faster than the rotation, i.e., there is  the variation over time of the orientation of the symmetry axis of the heavy fragment.

\subsection{Comparison with the numerical results}
\label{cnum}

Diagonalizing the molecular Hamiltonian \eqref{hamiltonian}and expanding  the spherical functions over the superposition of states with different $K$ values (\ref{Kproj1}-\ref{expansion}), we obtain
\begin{eqnarray}
\Psi_{nIM}&=&\sum_{K}\sqrt{\frac{2 I+1}{8\pi^2}}D^{I*}_{M, K}(\tilde \Omega) \phi_K(\epsilon), \nonumber \\
 \phi_K(\epsilon)&=&\sum_{L_H L_R}\sqrt{\frac{2L_H+1}{2I+1  }}a_{L_H L_R}C^{I K}_{L_H 0 L_R K}Y_{L_R K}(\epsilon, 0).
    \label{expansion1}
\end{eqnarray}

The probability distribution of various $K$ values in the wave function $\Psi_{nIM}$ is given by the following quantity
\begin{eqnarray}
P_{nI}(K)=\langle\phi_K|\phi_K \rangle.
    \label{Kdist}
\end{eqnarray}

It is instructive to investigate the distribution $K$ values for different excited states of molecular system. As an illustrative example
we consider the nucleus $^{240}$Pu as a nuclear molecule that consists of $^{236}$U+$^{4}$He nuclei. The deformation of $^{236}$U is taken as $\beta=0.25$. At the rigid body limit of the moments of inertia  of the fragments,  we obtain $\Im_b$=7.1 $\hbar^2/$MeV and the stiffness parameter of the potential $C=5.7$ MeV. It results in
the critical value $K_{cr}\simeq 6$ (see Eq.(\ref{Kcr})).
\begin{figure}[t]
\includegraphics[bb=0 0 558 315,scale=0.55]{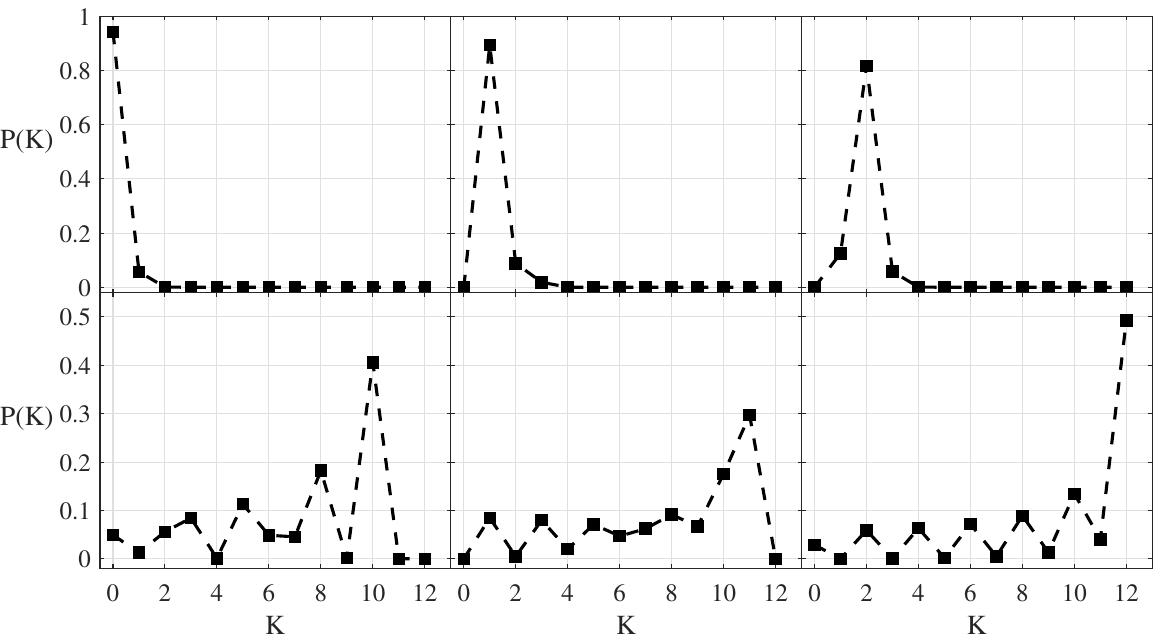}
\vspace{-.5cm} \caption{The probability distribution of $K$ values in various excited states of the molecular system at $I=12$. Top and bottom panels display the results for the pole-to-pole and
the equatorial configurations, respectively.}
\label{num2}
\end{figure}

\begin{figure}[t]
\includegraphics[scale=0.6]{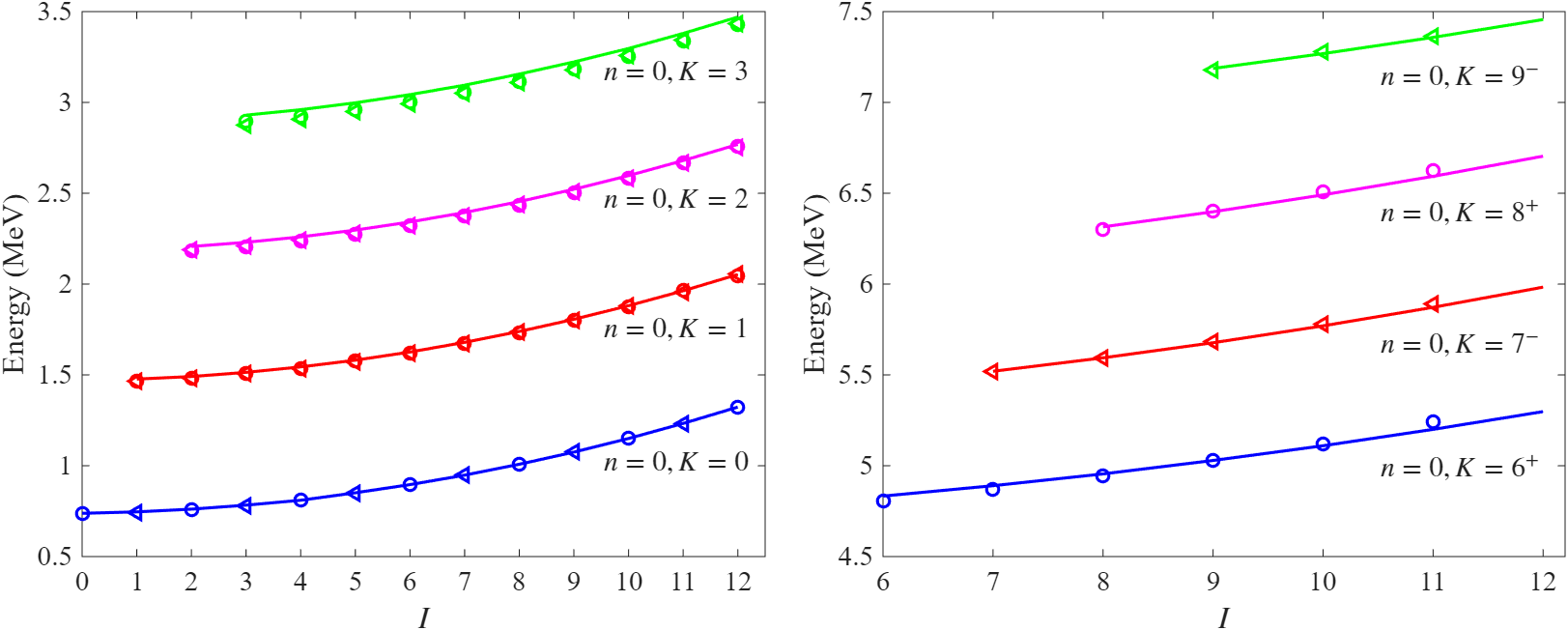}
\vspace{-.5cm} \caption{Energy spectra of rotational bands of the nucleus $^{240}$Pu, considered as a nuclear molecule that consists of $^{236}$U+$^{4}$He nuclei.
Left panel: the pole-to-pole configuration.
Right panel: the equator configuration.
Even (odd) spin states, obtained with aid of Eq. (\ref{ap13}) and Eq.(\ref{entot2}), are denoted by empty circles (empty triangles). The results, obtained by means  of numerical diagonalization of the Hamiltonian (\ref{fullH}) (see Sec.\ref{eig}), are connected by solid line.
}
\label{num22}
\end{figure}

We recall that for $K=0$, the potential energy exhibits  minima around the poles at $\epsilon=0$ and $\pi$ (black line) (see Fig.~\ref{efpot}). According to the analytical solution \eqref{analyt_wf}, in this case the nuclear molecule possesses the pole- to-pole configuration.  For $K=K_{cr}$, the potential energy becomes flat for all values of $\epsilon$, and  the spherical fragment can  freely move on the surface of the deformed one. In this case, one would expect a strong mixture of the wave functions with various $K$ values. At $K>K_{cr}$ the new regime is developed, and the system prefers the equatorial configuration.

Indeed, the results of numerical calculations of the probability distribution of the wave functions with different $K$ confirms such a transition (see Figs.~\ref{num2}).  For the pole-to-pole configuration the  maxima of these probabilities are located  at $K=0,1,3$. In contrast, once the motion of the spherical fragment shifts to the rotation around the equatorial surface of the heavy fragment, the excited state maxima  are located  at $K=10,11,12$. It is notable to mention a remarkable agreement between the results of the full quantum-mechanical calculations
of the energy spectra (see Sect.\ref{eig}) and the analytical estimations for different configurations (see Fig.~ \ref{num22}).

In order to delve deeper into this result, by means of  Eq.~\eqref{expansion1} we split the wave functions into the components, corresponding to  different $K$ values. The components with a significant contribution to the total wave function are shown in Fig.~\ref{num34}.
\begin{figure}[t]
\includegraphics[bb= 0 0 422 347,scale=0.5]{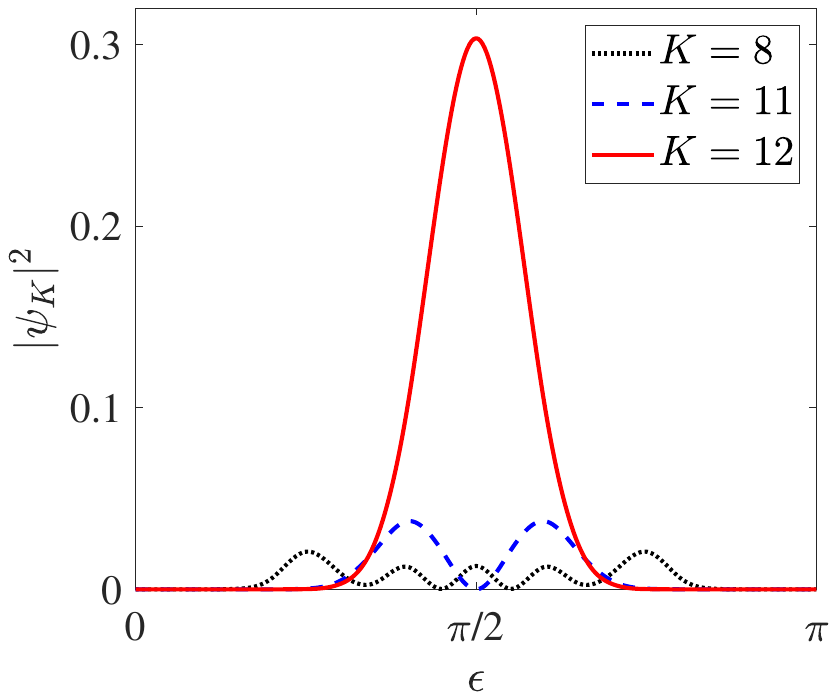}
\includegraphics[bb=0 0 422 347,scale=0.5]{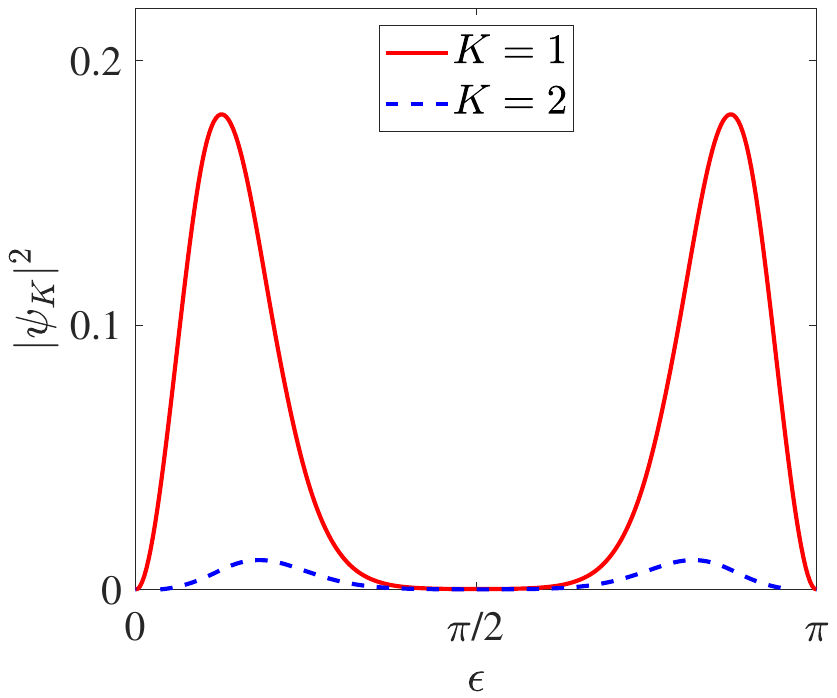}
\vspace{-.5cm} \caption{The probability density distribution of various components with different values of $K$: K=0 (black solid line);$K=K_{cr}$ (red dotted line), and $K>K_{cr}$ (blue dashed line).}
\label{num34}
\end{figure}
For yrast states, most of the wave functions are localized around $\epsilon=0$ and
$\epsilon=\pi$ (see the right panel). With the increase of the energy of excited states, the probability distributions of the wave functions with various $K$ values are distributed at all values of $\epsilon$. There are some states that behave in accordance with the equatorial limit. The probability distribution of these wave functions is displayed in the left panel, Fig.~\ref{num34}. The largest contribution to the wave function are given by the K=I component, while the  contribution of other components are suppressed. Moreover, that for this component, the values of $\epsilon$ are strongly bound around $\epsilon=\pi/2$ as expected from Eq.\eqref{approx_question}. We should emphasize, that for large $I$, only very limited amount of solutions exhibit such properties.

}

\section{Application to Experiment}
\label{desexp}

The formulated model can be used for various situations, when a nuclear system is considered as a nuclear molecule.
Below we present the description of excitation spectrum of a hyperdeformed  nucleus and angular anisotropy of the fission fragments, which are less discussed in literature,

\subsection{Hyperdeformation as a molecular state}
\label{4a}

As it was stressed in the Introduction, HD states  might be attributed to the formation of molecular states. Indeed, in certain mass regions it becomes very beneficial energetically to allocate a well-bound double magic cluster. To keep this benefit, other cluster(s) should also be well-bound or have a strongly deformed ground-state to reduce the Coulomb repulsion between clusters in a  molecule.
For example, the reflection-asymmetric HD states  can be identified with the third minimum in nuclei of the actinide mass region. In these nuclei, the fission barrier has a double-humped structure, leading to a local minimum characterized by the strong quadrupole and octupole deformations \cite{SGZhou_DHbarrier,Triple_Barr}. Experimentally, the states in the third minimum  are identified by studying the structure of the resonances, found in the reactions leading to the neutron-induced fission \cite{Kraszhoharkai}. The cluster systems, which define the structure of these extremely elongated reflection-asymmetric states, consist of strongly-bound fragment in the vicinity of double-magic $^{132}$Sn and the quadrupole-deformed fragment with $Z \approx 40$. Note, that the charge of the spherical fragment can deviates from $Z=50$ in order to balance the symmetry energy in the fragments of the molecular system. The well known example is the HD states in $^{232}$Th, whose properties are identical with the properties of $^{132}$Sn+$^{100}$Zr molecular system.


\begin{figure}[t]
\includegraphics[bb=0 0 158 522,scale=0.5]{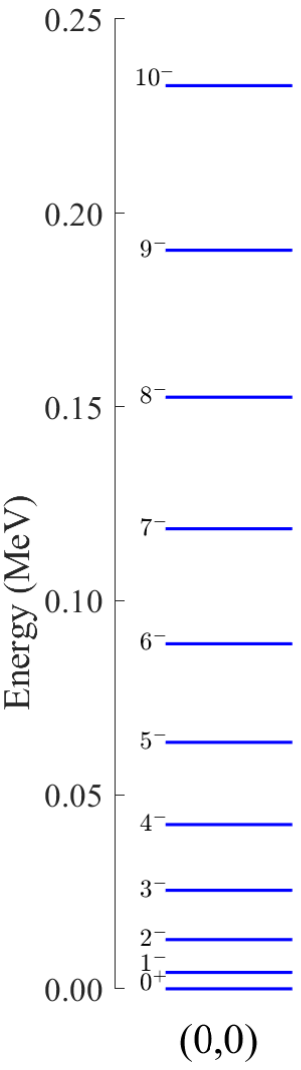}
\hspace{0.5cm}
\includegraphics[bb=0 0 269 522,scale=0.5]{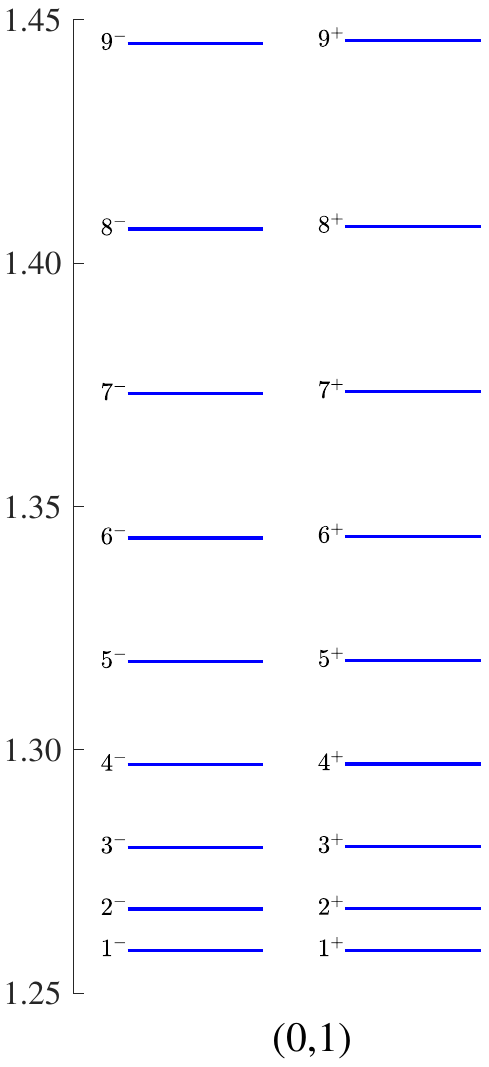}
\hspace{0.5cm}
\includegraphics[bb=0 0 307 522,scale=0.5]{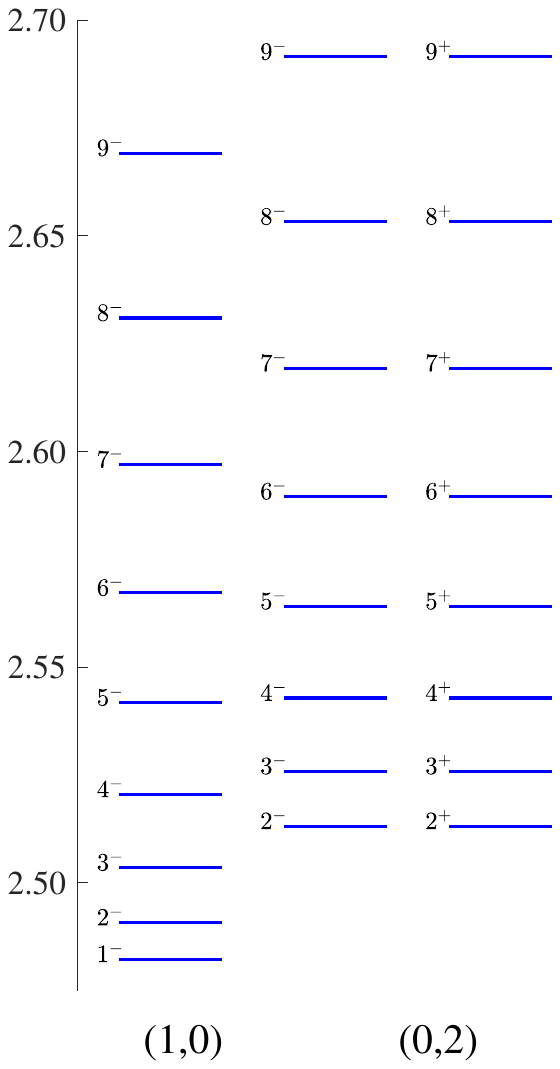}
\vspace{-.5cm} \caption{The spectrum of $^{132}$Sn+$^{100}$Zr molecular system, corresponding to the HD states in $^{232}$Th (see the text below). The following notation $(n,K)$  is used to characterize each band with the approximate quantum numbers. }
\label{num1}
\end{figure}

 The excitation spectrum of this molecule (see Fig.~\ref{num1}) have been calculated
 within our model (see Sec.\ref{eig})  by means of
 the numerical diagonalizing of the Hamiltonian \eqref{hamiltonian}.
The moments of inertia of $^{100}$Zr is calculated in the rigid body approximation
\begin{eqnarray}
\label{moiRB}
&&\Im^{r.b.}_{\mu' k'}=\frac{4\pi}{3}\rho_1\frac{R_1^5}{5}\int_{V_1} Y_{1 \mu'}^*(\theta',\phi')(1+\beta_{2}Y_{20}(\theta',\phi'))^5Y_{1 k'}(\theta',\phi')d\Omega\,.
\end{eqnarray}
Here, the quadrupole deformation parameter $\beta_2=0.35$ thar characterizes the deformation of $^{100}$Zr. The potential energy is calculated as a sum of the Coulomb and nuclear interactions, with nuclear interaction taken as a double-folding of fragments densities with effective nucleon-nucleon  interaction (Migdal forces) \cite{Adamian1996}. The calculated potential energy is fitted with the aid of the expression \eqref{potential}. Calculations yield $C(\beta_2,R_m(\epsilon))= 82$ MeV.

To reach a good convergency of the lowest energy levels it is enough to consider the basis  (\ref{bipol}) that includes states with
$L\leq 20 \hbar$.  Since the nucleus $^{100}$Zr is strongly deformed, the molecular system is tightly bound to the pole-to-pole configuration. As a result, the spectrum  consists of several bands, which can be classified in accordance with the expression \eqref{ap13}.
The considered system  is mass-asymmetric. This asymmetry leads to  considerable the reflection-asymmetric correlations that are exhibited themselves in the negligible parity splitting between positive and negative parity states.
The obtained results  are nicely reproduced by the approximate analytical solution \eqref{ap13} with the frequency $\hbar\omega_b \approx 0.59$MeV and the significant  moment of inertia $\Im \approx 250 \hbar^2$MeV$^{-1}$.

\begin{table}
    \centering
    \begin{tabular}{c|cccc}
        Nucleus & $^{232}$U  & $^{234}$U  & $^{236}$U & $^{238}$U\\
        \hline
        DNS & $^{94}$Sr+$^{138}$Xe  &  $^{96}$Sr+$^{138}$Xe &  $^{96}$Sr+$^{140}$Xe & $^{98}$Sr+$^{138}$Xe \\
        Energy (MeV) & 3.06   & 2.6 (3.1 $\pm$ 0.4) & 2.81 (2.7 $\pm$ 0.4) & 3.49 \\
         Rot. Const. (MeV) &1.83 &1.77 (2.1 $\pm$ 0.2)  & 1.75 (2.4 $\pm$ 0.4) & 1.70  \\
         $Q_2$ (10$^2$ e fm$^2$) & 92.4 & 93.0  & 93.47 & 96.77 \\
         $Q_3$ (10$^3$ e fm$^3$) & 30.0  & 28.5 & 29.9  & 27.8 \\
         \hline
    \end{tabular}
    \caption{The results of calculations of the excitation energy $E^*$ of the HD state,
    the rotational constant $1/(2\Im)$, the values of multipole moments
    $Q_2$ and  $Q_3$ for possible molecular systems.
    Number in parenthesis are experimental data:
    $^{234}$U (see Ref. \cite{Kraszhoharkai} $^{236}$U (see Ref. \cite{Thirolf}).
    }
    \label{tab1}
\end{table}
 In actinide regions several measurements of HD states are known experimentally \cite{Kraszhoharkai,Kraszhoharkai1,Thirolf}. The results of our calculations are in a good agreement with known experimental data for $^{234}$U, $^{236}$U (see Table \ref{tab1}). We provide as well the values of multipole moments $Q_2$ and  $Q_3$  that could be considered as our predictions, calculated for nuclear molecules.

\subsection{Angular anisotropy of the fission fragments}
\label{4b}

The molecular degrees of freedom play an important role in the fission process. It was shown,  that a fissile nucleus  can be considered as a system of two interacting clusters, i.e., as a nuclear molecule at the late stages of fission just before the separation of two primary fragments \cite{Andreev2005}.  This consideration have been used to calculate  mass, total kinetic energy and angular momentum distributions of fission fragments in neutron induced and spontaneous fission of actinides \cite{Andreev2005, Shneidman2002}. Here, we employ our model  to calculate the angular anisotropy of the fragments produced in neutron--induced fission of $^{239}$Pu.

Qualitatively, when  nucleus is moving along the fission barrier, the elongation  of the nuclear system increases. The increase of the elongation of the molecular configurations, which contribute to the nuclear wave function, can be achieved either by increase of the deformations of the fragments at touching configuration,  or by increase of  the mass asymmetry of the molecule itself.

The increase of elongation of the fissile nucleus up to the
saddle-point is achieved mainly by the increase of deformation of a mononucleus. The contribution of a molecular system wave function to the nuclear wave functions is negligible. After passing the  saddle point, however, the  increase of elongation is related mostly to  the rapid increase of the contribution of various binary systems, while the contribution of the mononucleus configuration vanishes.   Simultaneously, the molecule, characterized by the mass asymmetry $\xi$, has the  probability to decay relative to the    $R$, and, consequently, undergoing a fission.
The competition between the evolution and decay processes determines the dynamics of the fissioning  nucleus beyond the saddle point. Here, we consider the limiting case of  a well defined scission  configuration, assuming that the statistical equilibrium is established among various degrees of freedom of the system  before the  decay in $R$ (see also \cite{Andreev2005}).

Assuming that the decay of the scission-point
configuration is a fast process, the probability of emitting
fission fragments from the state $\Psi_{n,I,M=0, \pi}(\epsilon,\phi_R,\theta_R,\phi_H)$ in the direction defined by the angle
$\theta_R$ is given by:
\begin{eqnarray}
P_{n,K,I \pi}=2\pi\sin \theta_R  \int |\Psi_{n,I,M=0, \pi}|^2 \sin\epsilon d \epsilon d \phi_R d \phi_H \label{angdistr1}
\end{eqnarray}
Here, the change of the angular distribution caused by the Coulomb
excitation subsequent to the fission is neglected.

The angular distribution of fission fragments for the
channel characterized by  the angular momentum $I$, projection $M$,
parity $\pi$ at temperature $T$ can be written  as (see, e.g., \cite{huizenga}):
\begin{eqnarray}
W_{JM\Pi}(T,\theta)=N A(I)\sum_{n, K}\exp{\left [
-\frac{E_{n,J,M=0,\pi}}{T(E^{*})}\right ]}P_{n,K,{I \pi} }\,.
\label{angdistr5}
\end{eqnarray}
Here, the quantity $A(I)=\sqrt{2 I+1}$  is a geometric factor in a capture cross section for a realization of a particular $I$ channel, and $N$ is the normalization constant:
\begin{eqnarray}
N=   \left ( \sum_{n, K}\exp{\left [
-\frac{E_{n,J,M=0,\pi}}{T(E^{*})}\right ]} \right )^{-1}.
\end{eqnarray}
The energy of the particular state is defined by Eq.(\ref{ap13}).
To proceed further, we propose the following approximations.
First we assume that at the scission average masses of the fragments coincide with the maximum of experimental mass distribution. Second, we estimate deformations of the fragments by treating the average total kinetic energy as the energy of the molecular system at the top of the Coulomb barrier before the scission.
As a result, we obtain $(Z_1, A_1,\beta_1)$=(52, 132, 0.1) and
$(Z_2, A_2, \beta_2)$=(42, 108, 0.6). For these fragments calculations yield
$\Im_b$=250$\hbar^2$MeV$^{-1}$ and $\hbar \omega_b$=0.7 MeV. The moments
of inertia of the corresponding fragments are calculated in the rigid body limit. This choice is justified, since the DNS is highly excited at
the scission point. The temperature of the DNS is related to the excitation
energy $T(E^{*})=(E^*/a)^{1/2}$ with the level density parameter $a=A/12$ MeV$^{-1}$. The calculations, performed within the improved scission point model \cite{Rahmati2024}, yield $T \approx 0.64$ MeV.

\begin{figure}[t]
\begin{center}
\includegraphics[bb=30 13 541 399,scale=.5]{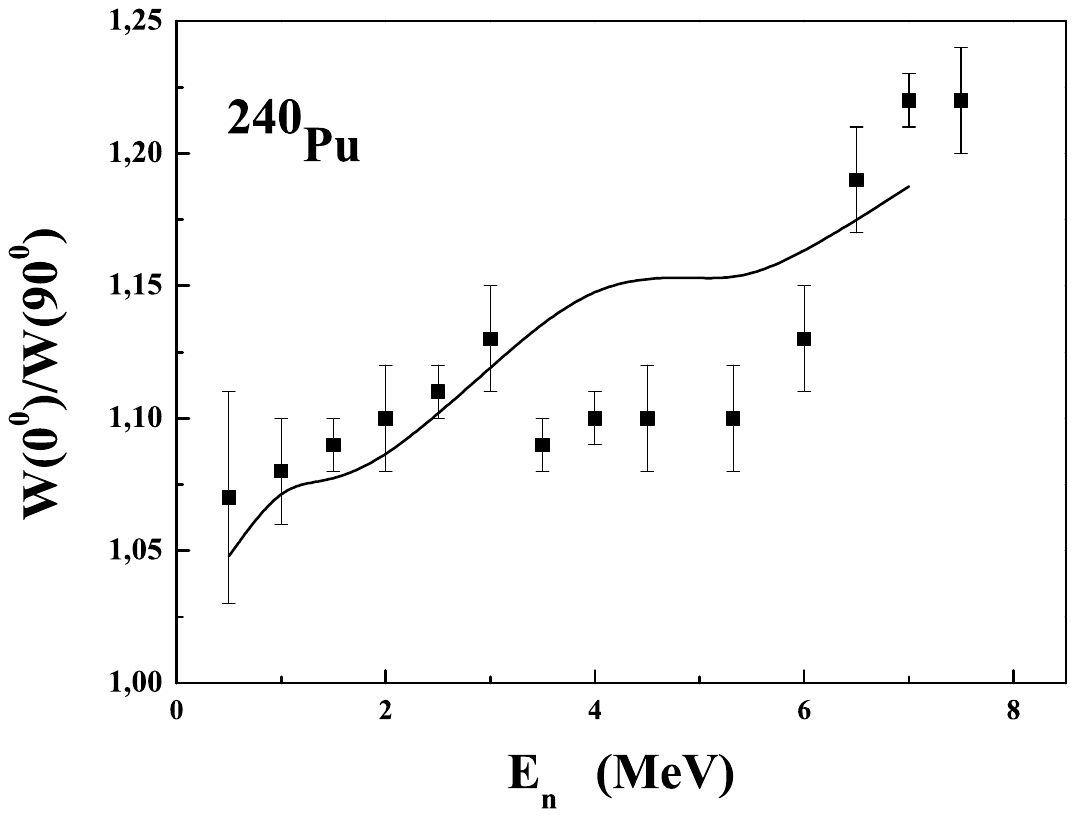}
\end{center}
\vspace{-.5cm} \caption{Calculated and experimental angular anisotropy
of fission fragments vs incident neutron energy. Experimental data are
taken from Ref.\cite{Simmons1965}.}
\label{anisotropy}
\end{figure}

Summing over the contributions of different channels, it is possible to compute the total angular distribution, $W(\theta)$. The quantity
that is commonly compared with experimental data is
the angular anisotropy, $W(\theta=0)/W(\theta=\pi/2)$, evaluated as a function of  $E_n$, as shown in Fig.~\ref{anisotropy}.
Our theoretical results are in a good agreement with experimental data at low energies ($E_n< 4$ MeV) . The behavior at higher energies, where the experimental angular anisotropy first decreases and then rapidly increases again above 6 MeV,
may be due to the interplay of two factors that are not yet included in our model. The decrease of the angular anisotropy may be related to the threshold of excitation
of  two-quasi-particle states in the fragments of the most favorable dinuclear configuration. The rapid increase above 6 MeV energy is  associated with the threshold of second-chance fission \cite{huizenga}. The amount of angular momentum carried away by the evaporated neutron is small compared with the angular momentum of the compound system. However, the excitation energy is significant  and, as a result, the angular anisotropy decreases.

\section{Summary}

We propose the theoretical approach (the model) for the description of the angular motion of two fragments in a molecular-like nuclear system. In this model one of the fragments is considered as a spherically symmetric, while another one is the axially deformed.
The nuclear interaction  plays the role of a "glue" between fragments of the molecule, which keeps the system in a touching configuration for a sufficiently long time. The dependence of the molecule potential upon the orientation in space is created by the Coulomb interaction.  The angular vibrational and rotational motions are responsible for the formation of low-energy excited states in such a molecular system.

The model enables a simple numerical diagonalization of the model Hamiltonian
in the basis of the bipolar spherical functions in the laboratory frame
(see Sect.\ref{sec:model}). In limiting cases  of small and large fragment deformations,
we have obtained the approximate analytical expressions
for the excitation spectrum and corresponding wave functions in the body fixed frame (see Sect.\ref{ansols}).
Such the approach provided the possibility to consider the coupling between the rotational and vibrational modes of the nuclear molecule in the second order of the perturbation theory.
As a result, contrary to the results presented in Refs. \cite{Uegaki1993,Uegaki2012} (focused on molecules  consisting of two light fragments), we have obtained the correct value of the moment of inertia of the nuclear molecule and analyzed the mixing of the $K$-values.
The proposed consideration becomes increasingly important in the case of molecules, consisting of heavy fragments.
Our analytical approach enabled to identify various types of the clusterization of two constituent fragments (nucleus) into a nuclear molecule. In the pole-pole configuration the rotational spectra of the molecule is complimented by two vibrational spectra of the spherical fragment relative to the pole of the heavy fragment (see Sect.\ref{an1}). In the equatorial configuration we found the manifestation of the precessional motion of the nuclear molecule around the total angular momentum of the system. Additionaly, the precessional motion is complemented by various types of the nutation depending on the mass contributions of the considered fragments (see Sect.\ref{an2}).

To illustrate the capabilities of the model  we have considered two cases that are described in significantly less details in literature (see Sect.\ref{desexp}).
As the first example, we have analyzed  the HD states in $^{232}$Th and  various U isotopes,
considered as nuclear molecules. In the case of the nucleus $^{232}$Th we have provided our predictions of the energy spectra of the HD states. A remarkable agreement between
the results of calculations  and available experimental data  was obtained for $^{232,234,236}$U.
As the second example, within our approach it was demonstrated that
angular anisotropy of fission fragments  depends strongly on the distribution of various $K$-values in the low-energy excitation spectrum of fissile nucleus at scission.  Applying the model, we were able to reproduce the experimental data on angular anisotropy in the neutron-induced fission of $^{240}$Pu.

The presented study of the excitation spectra of nuclear molecules is known to be important for various applications. In reactions with heavy ions, such studies allow to identify the appearance of resonances near the reaction threshold. In fission, a nucleus at scission can be presented as a superposition of various dinuclear systems. The angular motion in such systems is important for the generation of angular momenta of fission fragments and for their angular distribution. In addition, the collective vibrational and rotational states give rise to the collective enhancements of level densities of fissile nucleus, thus strongly affecting the distribution of various fission observables, including the TKE and neutron multiplicities. In nuclear structure, hyperdeformed nuclear states can also be considered as a manifestation of nuclear molecule properties.  Many of these phenomena are extensively discussed in the literature in different approximations.
We have developed a consistent approach that allows to analyze the above mentioned problems within the unified model.

 \appendix
\renewcommand{\theequation}{A-\arabic{equation}}
\setcounter{equation}{0}
\section{Time dependence of the coordinate {\bf R}}
\label{ap1}

In spherical coordinates for the time-dependence of the scalar
${\bf \dot{R}}^2$ in Eq.(\ref{ke5a})
we obtain  the following expression
\begin{eqnarray}
    {\bf \dot{R}}^2&=&\frac{4\pi}{3}{\dot R}^2 \sum_\mu |Y_{1\mu}(\theta_R,\phi_R)|^2 \nonumber \\
    &+& \frac{4\pi}{3}{\dot R}R \left (\sum_\mu \frac{d Y_{1\mu}(\theta_R,\phi_R)}{dt}Y^*_{1\mu}(\theta_R,\phi_R) +\sum_\mu Y_{1\mu}(\theta_R,\phi_R)\frac{d Y^*_{1\mu}(\theta_R,\phi_R)}{dt} \right ) \nonumber \\
    &+& \frac{4\pi}{3}{ R}^2 \sum_\mu|\frac{d Y_{1\mu}(\theta_R,\phi_R)}{dt}|^2\,.
    \end{eqnarray}
The sum in the first term is reduced to the form (see the textbook \cite{var}):
\begin{eqnarray}
\sum_\mu |Y_{1\mu}(\theta_R,\phi_R)|^2=\frac{3}{4\pi}\,.
\end{eqnarray}
The sums in the second term  are vanished:
\begin{eqnarray}
&&\sum_\mu \frac{d Y_{1\mu}(\theta_R,\phi_R)}{dt}Y^*_{1\mu}(\theta_R,\phi_R) +\sum_\mu Y_{1\mu}(\theta_R,\phi_R)\frac{d Y^*_{1\mu}(\theta_R,\phi_R)}{dt}  \nonumber \\
&&=\frac{d}{dt}\sum_\mu |Y_{1\mu}(\theta_R,\phi_R)|^2 = 0
\end{eqnarray}
To simplify the third term we consider the time derivatives of spherical functions:
\begin{eqnarray}
\frac{d Y_{1 \mu} (\theta_R,\phi_R)}{dt} =\frac{\partial Y_{1 \mu} (\theta_R,\phi_R)}{\partial\theta_R}\dot{\theta}_R+\frac{\partial Y_{1 \mu} (\theta_R,\phi_R)}{\partial \phi_R}\dot{\phi}_R\,.
\end{eqnarray}
Let us consider separately the contributions of the various factors that appear in the product
of the time derivatives
\begin{eqnarray}
&&\sum_\mu\frac{\partial Y_{1 \mu} (\theta_R,\phi_R) }{\partial \theta_R}
\frac{\partial Y^*_{1 \mu} (\theta_R,\phi_R) }{\partial \phi_R} =-i
\sum_\mu \mu \frac{\partial Y_{1 \mu} (\theta_R,\phi_R) }{\partial \theta_R}Y^*_{1\mu} (\theta_R,\phi_R)\nonumber\\
&&=-\frac{i}{2}\frac{\partial }{\partial \theta_R}\sum_\mu \mu |Y_{1 \mu}(\theta_R,\phi_R)|^2=0,
\end{eqnarray}
\begin{eqnarray}
  && \sum_\mu |\frac{\partial Y_{1 \mu}(\theta_R,\phi_R) }{\partial\theta_R}|^2 =
  \frac{\partial}{\partial \theta_R} \left (\sum_\mu Y_{1\mu}(\theta_R,\phi_R)\frac{\partial Y^*_{1\mu}(\theta_R,\phi_R)}{\partial \theta_R} \right )\nonumber \\
  && -\sum_\mu Y_{1\mu}(\theta_R,\phi_R)\frac{\partial^2 Y^*_{1\mu}(\theta_R,\phi_R)}{\partial \theta_R^2}
  =\sum_\mu Y_{1\mu}(\theta_R,\phi_R) \left (2-\frac{\mu^2}{\sin^2 \theta_R} \right )Y^*_{1\mu}(\theta_R,\phi_R)\nonumber \\
  &&+\cot \theta_R \sum_m Y_{1\mu}(\theta_R,\phi_R)\frac{\partial}{\partial \theta_R}Y^*_{1\mu}(\theta_R,\phi_R)
  =\frac{6}{4\pi}-\frac{1}{\sin^2 \theta_R}\sum_\mu \mu^2 |Y_{1\mu}(\theta_R,\phi_R)|^2.
\end{eqnarray}
Using the relation
\begin{eqnarray}
\sum_\mu \mu^2 |Y_{1\mu}(\theta_R,\phi_R)|^2=\frac{3}{4\pi}\sin^2\theta_R\,,
\label{mu2}
\end{eqnarray}
we obtain
\begin{eqnarray}
     \sum_\mu |\frac{\partial Y_{1 \mu}(\theta_R,\phi_R) }{\partial\theta_R}|^2=\frac{3}{4\pi}.
\end{eqnarray}
Expression (\ref{mu2}) can also be used to obtain the following expression:
\begin{eqnarray}
    \sum_\mu |\frac{\partial Y_{1 \mu} (\theta_R,\phi_R)}{\partial\phi_R}|^2=\frac{3}{4\pi} \sin^2 \theta_R.
\end{eqnarray}
Taking into account the above results, we obtain:
\begin{eqnarray}
\frac{4\pi}{3}\sum_\mu |\frac{d Y_{1\mu}(\theta_R,\phi_R)}{dt}|^2=\dot {\theta}_R^2+\sin^2\theta_R \dot {\phi}_R^2.
\label{derivative}
\end{eqnarray}

\renewcommand{\theequation}{B-\arabic{equation}}
\setcounter{equation}{0}

\section{Time-dependence of the coordinate $\mathbf{r_1}$}
\label{dfrag}

To carry out the task we transform the vector $\mathbf{r_1}$ with the aid of the Euler angles
$\Omega_H=\{\phi_H,\theta_H,0\}$ from the laboratory frame
to the intrinsic frame of the first fragment:
\beq
(r_1)_{1\mu}=\sqrt{\frac{4\pi}{3}}r_1Y_{1\mu}(\theta,\phi)=\sqrt{\frac{4\pi}{3}}r_1 \sum_{\mu'}{D}^{1*}_{\mu \mu'}(\Omega_H)Y_{1\mu'}(\theta',\phi')\,.
\eeq
Subsequently, taking into account that:
i)$\rho_1({\bf r_1})\equiv\rho_1$;
ii) the time dependence of the coordinate $r_1$ is determined by the
collective rotation of the deformed fragment; we have
\begin{eqnarray}
\langle {\bf \dot { r}_1}^2\rangle=\int_{V_1} \rho_1 {\bf \dot { r}_1}^2 d{\bf r_1}&=&\frac{4\pi}{3} \sum_{\mu' k'}\sum_\mu \dot{D}^1_{\mu \mu'}(\Omega_H)\dot{D}^{1*}_{\mu k'}(\Omega_H) \int_{V_1} \rho_1 r^2_1 Y_{1 \mu'}^*(\theta',\phi')
Y_{1 k'}(\theta',\phi')d{\bf r_1} \nonumber \\
&=& \sum_{\mu' k'}\sum_\mu \dot{D}^1_{\mu \mu'}(\Omega_H)\dot{D}^{1*}_{\mu k'}(\Omega_H)\Im_{\mu' k'}\,.
\label{222}
\end{eqnarray}
Here, the quantity
\begin{eqnarray}
\label{moi}
&&\Im_{\mu' k'}=\frac{4\pi}{3}\int_{V_1} \rho_1 r^2_1 Y_{1 \mu'}^*(\theta',\phi')
Y_{1 k'}(\theta',\phi')d{\bf r_1} =
\Im_{(\mu')}\delta_{\mu' k'}
\end{eqnarray}
is associated with the moment of inertia of the axially-symmetric deformed fragment.

Since our goal is the quantized form of the kinetic energy, it is quite natural to
assume the absence of the rotation around the symmetry axis
for the axially-symmetric fragment. In other words, we consider:
\begin{equation}
\Im_{zz}\propto \langle x^2+y^2\rangle\Rightarrow 0,
\Im_{xx}=\Im_{yy}\propto \langle x^2+z^2 \rangle\Rightarrow \Im_H \delta_{\mu' 0}\delta_{k' 0}.
\end{equation}
On the other hand, in spherical coordinates
$\Im_{(\mu)}\propto \langle |r_\mu|^2 \rangle$ :
\begin{equation}
\label{double}
\Im_{(\mu=\pm1)}= \frac{\Im_{zz}}{2},\quad  \Im_{(0)}\propto \langle z^2 \rangle\Rightarrow \Im_{H}-\frac{\Im_{zz}}{2}\,.
\end{equation}
Consequently, taking into account the above discussion, we can establish the connection
between  the Cartesian and spherical coordinate representations of the second rank tensor:
\begin{equation}
\left(
\begin{array}{cc}
\Im_{xx}=\Im_{yy}=\Im_H\\
\Im_{zz}=0
\end{array}
\right)
\Rightarrow
\left(
\begin{array}{cc}
\Im_{(1)}=\Im_{(-1)}=0\\
 \Im_{(0)}=\Im_H
\end{array}
\right)
\end{equation}
As a result, in virtue of Eq.~(\ref{moi}) and the relation
$\Im_{\mu' k'}=\Im_H \delta_{\mu' 0}\delta_{k' 0}$
we have
\begin{eqnarray}
\label{rdot}
\langle {\bf \dot { r}_1}^2 \rangle & = & \int_{V_1} \rho_1 {\bf \dot { r}_1}^2 d{\bf r_1} = \Im_H \sum_\mu \dot{D}^1_{\mu 0}(\Omega_H)\dot{D}^{1*}_{\mu 0}(\Omega_H)\nonumber \\
&=& \frac{4\pi}{3}\Im_H \sum_\mu |\frac{d Y_{1\mu}(\theta_H,\phi_H)}{dt}|^2
= \Im_H (\dot {\theta}_H^2+\sin^2\theta_H \dot {\phi}_H^2)\,.
\end{eqnarray}

\renewcommand{\theequation}{C-\arabic{equation}}
\setcounter{equation}{0}
\section{Classical kinetic energy of the deformed fragment}
\label{qkinen}

In order to analyze the various asymptotic solutions of the angular motion in the
molecule, it is convenient to transform the kinetic energy in  Eq.~\eqref{ke5a}  into the body--fixed coordinate system.
Vectors ${\bf r_1}$ and ${\bf R}$ in Eq.~\eqref{ke5a} are
\begin{eqnarray}
{\bf r_1}&=&\sum_m r_m {\bf e^m}=\sqrt{\frac{4\pi}{3}}r_1\sum_{m}D^{1*}_{m}(\tilde{\Omega}), \nonumber \\
{\bf R}&=&\sum_m R_m {\bf e^m}=\sum_{m,m'}D^{1*}_{m m'}(\tilde{\Omega}) R^{'}_{m^{'}} {\bf e^m}=\sqrt{\frac{4\pi}{3}}R\sum_{m,m',m''}D^{1*}_{m\ m'}(\tilde{\Omega})D^{1*}_{m^{'} m{''}}(\Omega_\epsilon) {\bf e^m}.
\end{eqnarray}

Consequently, the kinetic energy in \eqref{ke5a} takes the form:
\begin{eqnarray}
T&=&\ \frac{\Im_H}{2} \sum_m  \dot{D}^1_{m 0}(\tilde \Omega)\dot{D}^{1*}_{m 0}(\tilde \Omega)\nonumber \\
&+&\frac{\Im_R}{2} \sum_{m,m',m''}  \dot{D}^1_{m m'}(\tilde \Omega)D^{1*}_{m m''}(\tilde \Omega)D^{1*}_{m' 0}(\Omega_\epsilon )\dot{D}^1_{m'' 0}(\Omega_\epsilon)\nonumber \\
&+&\frac{\Im_R}{2} \sum_{m,m',m''}  D^1_{m m'}(\tilde \Omega)\dot{D}^{1*}_{m m''}(\tilde \Omega)\dot{D}^{1*}_{ m' 0}(\Omega_\epsilon )D^1_{m'' 0}(\Omega_\epsilon)\nonumber \\
&+&\frac{\Im_R}{2} \sum_{m,m',m''}  \dot{D}^1_{m m'}(\tilde \Omega)\dot{D}^{1*}_{m m''}(\tilde \Omega)D^{1*}_{ m' 0}(\Omega_\epsilon )D^1_{m'' 0}(\Omega_\epsilon)\nonumber \\
&+& \frac{\Im_R}{2} \sum_m  \dot{D}^1_{m 0}( \Omega_\epsilon)\dot{D}^{1*}_{m 0}( \Omega_\epsilon).
\label{class_kinetic}
\end{eqnarray}
Taking the derivatives of Wigner functions and simplifying the expression, we obtain the following expression for the kinetic energy
\begin{equation}
T=\frac{1}{2}\sum_{i, j =1}^4 \hat{G}_{i j} \dot{\theta}_i\dot{\theta}_j\,.
\end{equation}
where  $\theta_1=\phi_H, \theta_2=\theta_H, \theta_3=\phi_R, \theta_4=\epsilon$.

The matrix elements have the following form:
\begin{eqnarray}
G_{11}&=& (\Im_H+\Im_R)\sin^2{\theta_2}+\Im_R \sin{2\theta_2}\cos{\theta_3}\sin{\theta_4}\cos{\theta_4}+\Im_R(\cos^2{\theta_2}-\cos^2{\theta_3}\sin^2\theta_2)\sin^2\theta_4, \nonumber \\
G_{12}&=&G_{21}=\Im_R( \sin^2 \theta_4 \cos \theta_3 \sin \theta_3 \sin \theta_2- \sin \theta_4 \cos \theta_2 \sin \theta_3), \nonumber \\
G_{13}&=&G_{31}=\Im_R (\sin^2 \theta_4 \cos \theta_2+ \sin \theta_4 \cos{\theta_4} \cos \theta_3 \sin \theta_2), \nonumber \\
G_{14}&=& G_{41}= \Im_R \sin \theta_2 \sin \theta_3, \nonumber \\
\label{gf2}
G_{22}&=&(\Im_H+\Im_R)-\Im_R \sin^2 \theta_4 \sin^2 \theta_3, \\
G_{23}&=&G_{32}=-\Im_R \sin\theta_3 \sin \theta_4 \cos{\theta_4}, \nonumber \\
G_{24}&=&G_{42}=\Im_R \cos \theta_3, \nonumber \\
G_{33}&=&\Im_R \sin^2 \theta_4, \nonumber \\
G_{34}&=&G_{43}=0, \nonumber \\
G_{44}&=&\Im_R, \nonumber\\
G_{55}&=& \mu \nonumber\,.
\end{eqnarray}

The determinant of the matrix $G$ has a surprisingly simple form:
\begin{eqnarray}
G=\textrm{Det}(\hat{G})=\Im_R^2 \Im_H^2 \sin^2{\theta_2}\sin^2{\theta_4}.
\end{eqnarray}

For the inverse matrix $G^{-1}\equiv A$ we have:
{\begin{eqnarray}
\label{eqa2}
&&A_{11} = \frac{\csc^2\theta_2}{\Im_{H}},\quad
A_{12} =A_{21}=0, \quad
A_{13} = A_{31}=- \frac{(\cot \theta_2 +\cos\theta_3 \cot\theta_4)\csc\theta_2}{\Im_{H}}, \nonumber\\
&&A_{14} = A_{41} = -\frac{\csc\theta_2\sin\theta_3}{\Im_{H}}, \quad
A_{22} = \frac{1}{\Im_{H}}, \quad
A_{23} = A_{32} = \frac{\cot\theta_4\sin \theta_3}{\Im_{H}}, \\
&&A_{24} = A_{42} = - \frac{\cos\theta_3}{\Im_{H}}, \quad
A_{33} = \frac{\Im_{H}\left[\cot^2\theta_2-1+2\cos\gamma\cot\theta_2\cot\theta_4\right]+(\Im_{H}+\Im_{R})\csc^2\theta_4}{\Im_{H}\Im_{R}}, \nonumber\\
&&A_{34} = A_{43} = \frac{\cot \theta_2 \sin \theta_3}{\Im_{H}},\quad
A_{44} = \frac{1}{\Im_{b}}, \quad A_{55} = 1/\mu \nonumber \,.
\end{eqnarray}
Here, we introduce the definition
\beq
\label{invmb}
\frac{1}{\Im_{b}}=\frac{1}{\Im_{R}}+\frac{1}{\Im_{H}}\,.
\eeq

\renewcommand{\theequation}{D-\arabic{equation}}
\setcounter{equation}{0}
\section{The Coulomb potential of a nuclear molecule}
\label{calc3}
\begin{figure}[bth]
\includegraphics[bb=0 0 506 371,scale=.7]{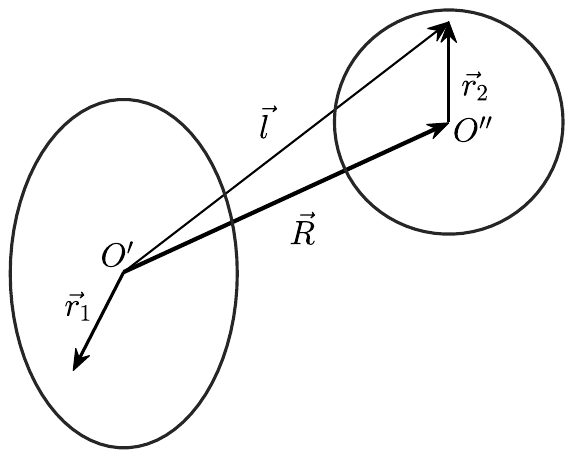}
\vspace{-.3in}
\caption{Sketch of points in different volume elements of the molecule}
\label{fig7}
\end{figure}

Our aim is to evaluate the Coulomb energy between axially-symmetric and spherical fragments in a
nuclear molecule (see Fig.\ref{fig7}). The Coulomb potential of two fragments (nuclei) in the molecule can be written in the following form
\begin{eqnarray}
U_C=e^2\int \frac{\rho_1({\bf r_1})\rho_1({\bf r_2})}{|{\bf r_2}-{\bf r_1}+{\bf R}|}d{\bf r_1}d{\bf r_2}
\label{Clmb1}
\end{eqnarray}

Here, the vectors ${\bf r_1}$ and ${\bf r_2}$ are measured from the centers of fragments "1" and "2" towards the corresponding points in volume elements, respectively. The vector ${\bf R}$ connects the centers of the two fragments. Next, we introduce the vector ${\bf l}$ that connects the point in the fragment "2" with the center of the first fragment. Evidently, between the vector lengths of ${\bf l}$ and ${\bf r_1}$ there is the inequality
$ l\geqslant r_1$. Taking into account this fact and the relation ${\bf l}={\bf r_2}-(-{\bf R})$, we obtain (see the textbook \cite{var})
\begin{eqnarray}
\frac{1}{|{\bf l}-{\bf r_1}|}&=&\sum_{\lambda_1}\frac{4\pi}{2 \lambda_1+1}\frac{r_1^{\lambda_1}}{l^\lambda_1+1}\left (Y_{\lambda_1}(\Omega_1)\cdot Y_{\lambda_1}(\Omega_{{\bf l}}) \right ) \nonumber \\
    \label{Clmb2}
&=&\sum_{\lambda_1 \mu_1}(-1)^{\mu_1}\frac{4\pi}{2 \lambda_1+1}\frac{r_1^{\lambda_1}}{l^\lambda_1+1} \hat Q^{(1)}_{\lambda_1 -\mu_1} \frac{ Y_{\lambda_1 \mu_1}(\Omega_{{\bf l}})}{l^{\lambda_1+1}}\,,
\end{eqnarray}
where
$ \hat Q^{(1)}_{\lambda_1 \mu_1}=r_1^{\lambda_1} Y_{\lambda_1 \mu_1}(\Omega_1)$
is a multipole operator in the deformed fragment "1":

Bearing in mind $r_2 \leqslant R$,
we have
\begin{eqnarray}
\frac{Y_{\lambda_1 -\mu_1}(\Omega_{\bf l})}{l^{\lambda+1}}&=&\sum_{\lambda_2}\sqrt{\frac{4\pi (2\lambda_1+2\lambda_2)!}{(2\lambda_1)!(2\lambda_2+1)!}}\frac{r_2^{\lambda_2}}{R^{\lambda_1+\lambda_2+1}} \left \{ Y_{\lambda_2}(\Omega_2) \times Y_{\lambda_1+\lambda_2+1}(\Omega_{-{\bf R}})\right \}_{\lambda_1 -\mu_1} \nonumber \\
 \label{Clmb4}
&=&\sum_{\lambda_2}\sum_{\mu_2 \mu}C^{\lambda_1 -\mu_1}_{\lambda_2 \mu_2, \lambda_1+\lambda_2 \mu}\sqrt{\frac{4\pi(2\lambda_1+2\lambda_2)! }{(2\lambda_1)!(2\lambda_2+1)!}}
\hat Q^{(2)}_{\lambda_2 \mu_2}
(\Omega_2)\frac{Y_{\lambda_1+\lambda_2,\mu}(\Omega_{-\bf{R}})}{R^{\lambda_1+\lambda_2+1}}\,,
\end{eqnarray}
where $\hat Q^{(2)}_{\lambda_2 \mu_2}=r_2^{\lambda_2} Y_{\lambda_2 \mu_2}(\Omega_2)$ is
the multipole operator in the second, spherical fragment.
Using the relation
\begin{eqnarray}
Y_{\lambda_1+\lambda_2,\mu}(\Omega_{-\bf{R}})=(-1)^{\lambda_1+\lambda_2}Y_{\lambda_1+\lambda_2,\mu}(\Omega_{\bf{R}})\,,
    \label{Clmb6}
\end{eqnarray}
we transform Eq.(\ref{Clmb4}) to the form
\begin{eqnarray}
&&\frac{Y_{\lambda_1 -\mu_1}(\Omega_{\bf l})}{l^{\lambda+1}}=\sum_{\lambda_2}\sum_{\mu_2 \mu} (-1)^{\lambda_1+\mu_2}C^{\lambda_1+\lambda_2, -\mu}_{\lambda_1 \mu_1, \lambda_2 \mu_2}\nonumber \\
&&\times\sqrt{\frac{2\lambda_1+1}{2\lambda_1+2\lambda_2+1}}\sqrt{\frac{4\pi(2\lambda_1+2\lambda_2)! }{(2\lambda_1)!(2\lambda_2+1)!}}\hat Q^{(2)}_{\lambda_2 \mu_2}\frac{Y_{\lambda_1+\lambda_2,\mu}(\Omega_{\bf{R}})}{R^{\lambda_1+\lambda_2+1}}.
    \label{Clmb7}
\end{eqnarray}
By means of Eqs.~\eqref{Clmb2} and \eqref{Clmb7}, we obtain
for the definition of the Coulomb energy of the molecule (\ref{Clmb1})
the following expression
\begin{eqnarray}
U_C&=&e^2\sum_{\lambda_1 \lambda_2}(-1)^{\lambda_1+\mu}\frac{4\pi}{2\lambda_1+2\lambda_2+1}\sqrt{\frac{4\pi(2\lambda_1+2\lambda_2+1)! }{(2\lambda_1+1)!(2\lambda_2+1)!}}  \nonumber \\
&\times&\left \{Q^{(1)}_{\lambda_1}\times
Q^{(2)}_{\lambda_2}\right \}_{(\lambda, -\mu)} \frac{Y_{\lambda_1+\lambda_2}(\Omega_{{\bf R}})}{R^{\lambda_1+\lambda_2+1}},
    \label{Clmb8}
\end{eqnarray}
where
\begin{eqnarray}
Q^{(i)}_{\lambda \mu}=\int \rho_i({\bf r})r^\lambda Y_{\lambda \mu}(\Omega)d{\bf r}\,, \quad i=1,2\,,
    \label{Clmb9}
\end{eqnarray}
are the multipole moments of the first and the second fragments.
For the first, axially-symmetric fragment we have:
\begin{eqnarray}
Q^{(1)}_{\lambda_1 \mu_1}&=&\sum_{\mu_1'}D^{\lambda_1*}_{\mu_1 \mu_1'}(\Omega_H)  \int \rho_i({\bf r})r^\lambda Y_{\lambda \mu}(\Omega')d{\bf r}=
Y_{\lambda_1 \mu_1}(\Omega_H)Q^{(H)}_{\lambda_1}(\beta_2)\delta_{\mu_1 0},
\nonumber \\
Q^{(H)}_{\lambda_1}(\beta_2)&=&\sqrt{\frac{4\pi}{2\lambda_1+1}}\int \rho_i({\bf r})r^\lambda Y_{\lambda 0}(\Omega')d{\bf r}.
    \label{Clmb10}
\end{eqnarray}
For the spherical fragment "2", we have the multipole moment
\begin{eqnarray}
Q^{(2)}_{\lambda_2 \mu_2}=\frac{Z_2}{\sqrt{4\pi}}\delta_{\lambda_2 0} \delta_{\mu_2 0}.
    \label{Clmb11}
\end{eqnarray}
Taking into account Eqs.(\ref{Clmb8},\ref{Clmb10},\ref{Clmb11}), we obtain  the following expression for the Coulomb energy
\begin{eqnarray}
V_C=\sum_{\lambda} \frac{e^2 Z_2 Q^{(H)}_{\lambda}(\beta_{2})} {R^{\lambda+1}}P_\lambda (\cos{\varepsilon})
    \label{Clmb12}
\end{eqnarray}
between the axially-symmetric and spherical fragments (nuclei).

In linear order with respect to the deformation parameter $\beta_2$ we obtain the well known Wong formula \cite{Wong}:
\begin{eqnarray}
V_C=e^2\frac{Z_1 Z_2}{R}+\sqrt{\frac{9}{20\pi}}e^2\frac{Z_1 Z_2}{R^3} R_1^2 \beta_2 P_2(\cos{\epsilon})\,.
    \label{wong}
\end{eqnarray}

\renewcommand{\theequation}{E-\arabic{equation}}
\setcounter{equation}{0}
\section{The evaluation of the interaction
$V_{int}$ in the perturbation approach}
\label{calc2}


\begin{widetext}
We start with the following matrix elements between rotational wave functions
\begin{eqnarray}
&&\langle I M K_1 \pi_1|\hat I_1 \hat I_3+\hat I_3 \hat I_1| I M K_2 \pi_2\rangle
=\frac{K_1+K_2}{2}
\left \{ \delta_{K_1,K_2-1}\sqrt{I(I+1)-K_2(K_2-1)} \right.\nonumber \\
&&+\left. \delta_{K_1,K_2+1}\sqrt{I(I+1)-K_2(K_2+1)}\right \}\delta_{\pi_1 \pi_2},
\hspace{20pt} (K_1 \ne 0, K_2 \ne 0), \nonumber \\
&&\langle I M K_1 \pi_1|\hat I_1 \hat I_3+\hat I_3 \hat I_1| I M K_2\pi_2\rangle
=\frac{\sqrt{I(I+1)}}{2} \left \{ \delta_{K_1 0}\delta_{K_2 1}+\delta_{K_1 1}\delta_{K_2 0}\right \}\delta_{\pi_1 \pi_2},
\label{ap9}
\end{eqnarray}
and
\begin{eqnarray}
&&\langle I M K_1 \pi_1|2 i \hat I_2| I M K_2\pi_2\rangle \nonumber \\
=
&&\left \{\delta_{K_1,K_2-1}\sqrt{I(I+1)-K_2(K_2-1)}
-\delta_{K_1,K_2+1}\sqrt{I(I+1)-K_2(K_2+1)}\right \}\delta_{\pi_1 \pi_2}, \nonumber \\
&&\langle I M K_1 \pi_1|2 i \hat I_2| I M K_2\pi_2\rangle = 2\sqrt{I(I+1)}\{\delta_{K_10}\delta_{K_2 1}-\delta_{K_11}\delta_{K_20} \}
\label{ap10}
\end{eqnarray}
The vibrational matrix elements of the approximated Hamiltonian \eqref{ap66}
have the following forms
\begin{eqnarray}
&& (  \phi_{n_1 K}|\frac{1}{\epsilon}|\phi_{n_2 K-1})  =
\begin{cases}
\frac{1}{\epsilon_0}\sqrt{\frac{n_1!}{n_2!}\frac{(n_2+K-1)!}{(n_1+K)!}}, & n_2 \le n_1 \\
0, & n_2 > n_1
\end{cases} \nonumber \\
&& (  \phi_{n_1 K}|\frac{1}{\epsilon}|\phi_{n_2 K+1})  =
\begin{cases}
\frac{1}{\epsilon_0}\sqrt{\frac{n_2!}{n_1!}\frac{(n_1+K)!}{(n_2+K+1)!}}, & n_1 \le n_2 \\
0, &  n_1 > n_2,
\end{cases} \nonumber \\
\label{ap11}
\end{eqnarray}
and
\begin{eqnarray}
 (  \phi_{n_1 K}|\frac{1}{\sqrt{\epsilon}}\frac{\partial}{\partial \epsilon}\sqrt{\epsilon}|\phi_{n_2 K-1})&=&
 \left (K+\frac{1}{2} \right )(  \phi_{n_1 K}|\frac{1}{\epsilon}|\phi_{n_2 K-1}) \nonumber \\
 &&-\frac{1}{\epsilon_0}\sqrt{n_1+K}\delta_{n_1 n_2}+\frac{1}{\epsilon_0} (\sqrt{n_1+1}-2\sqrt{n_2})\delta_{n_1,n_2-1}, \nonumber \\
(  \phi_{n_1 K}|\frac{1}{\sqrt{\epsilon}}\frac{\partial}{\partial \epsilon}\sqrt{\epsilon}|\phi_{n_2 K+1})&=&
- \left (K+\frac{1}{2} \right )(  \phi_{n_1 K}|\frac{1}{\epsilon}|\phi_{n_2 K+1})
\label{ap12} \\
&-&\frac{1}{\epsilon_0}\sqrt{n_2+K+1}\delta_{n_1 n_2}-\frac{1}{\epsilon_0} (\sqrt{n_2+1}-2\sqrt{n_1})\delta_{n_1,n_2-1}\,, \nonumber \\
\label{ap12a}
(  \phi_{n_1 K}|{\epsilon}\frac{\partial}{\partial \epsilon}|\phi_{n_2 K})&=& \\
&-&\delta_{n_1,n_2}+\sqrt{n_1(n_1+K)}\delta_{n_2,n_1-1}-\sqrt{n_2(n_2+K)}\delta_{n_2,n_1+1}\,.\nonumber
\end{eqnarray}
\end{widetext}
All matrix elements in  Eqs.(\ref{ap9})-(\ref{ap12}) are proportional to $\sim 1/\epsilon_0$. The major contribution of the matrix element (\ref{ap12a}) is defined by the diagonal first term.
The differences between the energies, corresponding to the different values of $n$ and $K$, are:
\begin{eqnarray}
\label{enggg}
 E_{n_1 K_1}-E_{n_2,K_2} \sim \hbar \omega = \frac{\hbar^2}{\Im_b \epsilon_0^2}\,,
\end{eqnarray}
Consequently, we have
\begin{eqnarray}
\label{4order}
E_{n_i}=E_{n_i}^{(0)}+\sum_{j\ne i}\frac{|V_{ij}|^2}{\omega_{ij}}
\end{eqnarray}
Here we have introduced the notations:
\begin{eqnarray}
&&n \Rightarrow\{IMKn\}, \quad E^{(0)}_n \Rightarrow E_{IMKn},\quad
\omega_{i j} \Rightarrow E_{I_{i} M_{i} K_{i} n_{i}}-E_{{I_j} M_{j} K_{j} n_{j}}\,,  \nonumber \\
&& V_{i j} \Rightarrow \langle \Phi^\pi_{I_i M_i K_i n_i}|V_{int}|\Phi^\pi_{I_j M_j K_j n_j}\rangle\,, 
\nonumber
\end{eqnarray}
where the functions $\Phi^\pi_{IMKn}$ are the basis states (\ref{WF_sym}).
By means of Eqs.(\ref{ap9})-(\ref{ap12}), we evaluate the contribution of
the interaction (\ref{enggg})
to the total energy:
\bea
 \sum_{k\ne n}\frac{|V_{nk}|^2}{\omega_{nk}}=-\frac{\hbar^2}{2\Im_H}\frac{\Im_R} {(\Im_H+\Im_R)}\left[I(I+1)-2K^2\right]\,.
\eea
To apply the perturbation theory to the Hamiltonian \eqref{app66d}, we need to evaluate another set of matrix elements for the vibrational degree of freedom $\tilde \epsilon$:

\begin{eqnarray}
    \label{vb1}
    (  \phi_{n_1 }|\tilde \epsilon|\phi_{n_2})  =
\frac{\tilde\epsilon_0}{\sqrt{2}}(\sqrt{n_1} \delta_{n_1,n_2+1}+\sqrt{n_2} \delta_{n_1,n_2-1})
\end{eqnarray}
and
\begin{eqnarray}
    \label{vb2}
(  \phi_{n_1 }| \frac{\partial}{\partial \tilde\epsilon} |\phi_{n_2})=
\frac{1}{\sqrt{2}\tilde \epsilon_0}\left (-\sqrt{n_1} \delta_{n_1,n_2+1}+ \sqrt{n_2}\delta_{n_1,n_2-1} \right ).
\end{eqnarray}



\begin{thebibliography}{99}

\bibitem{Bethe1936}
                  H. A. Bethe, R. F. Bacher,
                  Rev. Mod. Phys. {\bf 8}, 82 (1936).

\bibitem{Shneidman2003}
                 T.M. Shneidman, G.G. Adamian, N.V. Antonenko, R.V. Jolos, W. Scheid,
                 Phys. Rev. C {\bf 67}, 014313 (2003).

\bibitem{Shneidman2011}
                T.M. Shneidman, G.G. Adamian, N.V. Antonenko,
                R.V. Jolos,  W. Scheid,
                Eur.Phys.J. A {\bf 47}, 34 (2011).

\bibitem{Shneidman2015}
               T.M. Shneidman, G.G. Adamian, N.V. Antonenko, R.V. Jolos, Shan-Gui Zhou,
               Phys. Rev. C {\bf 92}, 034302 (2015).
\bibitem{Tanaka2021}
                 J. Tanaka, Z. Yang, S. Typel, et al.,
                 Science, {\bf 371}, 260 (2021).
\bibitem{Yamaya1990}
                 T. Yamaya, S. Oh-ami, M. Fujiwara, et al.,
                 Phys. Rev. C {\bf 42}, 1935 (1990).
\bibitem{Fukada2009}
                M. Fukada, M.K. Takimoto, K. Ogino, S. Ohkubo,
                Phys. Rev. C {\bf 80}, 064613 (2009).
\bibitem{Issatayev2023}
                T. Issatayev, N.K. Skobelev, T.M. Shneidman, Yu. E. Penionzhkevich, V. Burjan, J. Mr\'{a}zek,
                Phys.Part. and Nucl.Lett. 20, 988 (2023).

\bibitem{Uegaki1993}
                E. Uegaki, Y. Abe,
                Prog. Theor. Phys. {\bf 90}, 615 (1993).

\bibitem{Uegaki2012}
               E. Uegaki, Y. Abe,
               Prog. Theor. Phys. {\bf 127}, 831 (2012).

\bibitem{Volkov1982}
            V.V. Volkov, Phys.Pep. {\bf 44}, 93 (1978).
\bibitem{Antonenko}
            V.V. Volkov, G.G. Adamian, N.V. Antonenko, E.A. Cherepanov,
            A. K. Nasirov,
            Il Nuovo Cimento A {\bf 110}, 1127 (1997).

\bibitem{Andreev2005}
            A.V. Andreev, G.G. Adamian, N.V. Antonenko, S.P. Ivanova,
            Eur.Phys.J. A {\bf 26}, 327 (2005).

\bibitem{Rahmati2024}
        A. Rahmatinejad, A. V. Andreev, A. N. Bezbakh, A. V. Isaev, R. S. Mukhin and T. M. Shneidman,
        Int. J. Mod. Phys. E 2441018 (2024).

\bibitem{Wilkins1976}
           B.D. Wilkins, E.P. Steinberg, R.R Chasman,
           Phys. Rev. C {\bf 14}, 1832 (1976).


\bibitem{Nix1965}
            J.R. Nix, W.J. Swiatecki,
            Nucl. Phys. A {\bf 71}, 1 (1965).

\bibitem{Hess1984}
            P.O. Hess, W. Greiner,
            Nuovo Cimento A {\bf 83}, 76 (1984)

\bibitem{H2}
            P.O. Hess, W. Greiner, and W.T. Pinkston,
            Phys. Rev. Lett. {\bf 53}, 1535 (1984).

\bibitem{Misicu1999}
            \c{S}. Mi\c{s}icu, A. S\u{a}ndulescu, G. M. Ter-Akopian, W. Greiner,
            Phys. Rev. C {\bf 60}, 034613, (1999).

\bibitem{Shneidman2002}
        T.M. Shneidman, G.G. Adamian, N.V. Antonenko,
        S.P. Ivanova, R.V. Jolos, W. Scheid,
        Phys. Rev. C {\bf 65}, 064302 (2002).

\bibitem{AR2024}
    T.M. Shneidman, A. Rahmatinejad, G.G. Adamian,   N.V. Antonenko, Phys.Rev. C {\bf 111}, 064621 (2025).

\bibitem{Randrup}
      T. Dossing, S. \AA berg, M. Albertsson, B. G. Carlsson, and J. Randrup,
      Phys. Rev. C, {\bf{109}}, 034615 (2024).


\bibitem{Scheid}
        W. Greiner, J.Y. Park, W. Scheid,
        {\it Nuclear Molecules}  World Scientific, Singapore, 1995.

\bibitem{Begtsson1981}
        T. Bengtsson, M.E. Faber, G. Leander, P. M\"{o}ller, M. Ploszajczak, I. Ragnarsson, S. \AA berg,
        Phys. Scrip.  {\bf 24}, 200 (1981).


\bibitem{Nazarewicz1992}
         W. Nazarewicz, J. Dobaczewski,
         Phys. Rev. Lett.  {\bf 68}, 154 (1992).

\bibitem{ra}
         R. Nazmitdinov, S. {\AA}berg,
         Phys. Lett. {\bf B 289}, 238 (1992).


\bibitem{Aberg1994}
        S. {\AA}berg, L.-O. J\"onsson,
         Z. Physik A - Hadrons and Nuclei,
         {\bf 349}, 205 (1994).

\bibitem{Cwiok1994}
        S. \'{C}wiok, W. Nazarewicz, J.X. Saladin, W. P{\l}\'{o}ciennik,
        A. Johnson,  Phys. Lett. B {\bf 322}, 304 (1994).

\bibitem{Jonsson1997}
        L.-O. J\"onsson,
        Prog. Part. Nucl. Phys. {\bf 38}, 99 (1997).

\bibitem{Afanasjev2018}
         A.V. Afanasjev, H. Abusara, S.E. Agbemava,
         Phys. Scrip.  {\bf 93}, 034002  (2018).

\bibitem{SAROV}
         V.L. Grigorenko, N.V. Antonenko, I.A. Artyukov, et al.,
         FizMat  {\bf 1}, 123 (2023).

\bibitem{ELINP}
         A.S. Cucoanes, M. Cuciuc, Y. Arai, K. Homma, Y. Nakamiya, O. Tesileanu,
        Rom. J. Phys. {\bf 69}, 401 (2024).

\bibitem{sev}
           A.P. Severyukhin, S. {\AA}berg, N.N. Arsenyev, R.G. Nazmitdinov,
            Phys. Rev. C {\bf 95}, 061305(R) (2017).

\bibitem{var}
            V.K. Khersonskii, A.N. Moskalev, D.A. Varshalovich,
            {\it Quantum Theory Of Angular Momentum},
            World Scientific Publishing Company,
            Singapore, 1988.

\bibitem{Pauli}
           W. Pauli, in Handbuch der Physik,
           vol. XXIV, p. 120;
           Springer Verlag, Berlin, 1933.

\bibitem{Migdal}
         A.B. Migdal,
         {\it Qualitative Methods in Quantum Theory },
         CRC Press, Taylor and Francis Group, London and New York, 2000.

\bibitem{Adamian1996}
         G.G. Adamian, N.V. Antonenko, R.V. Jolos, S.P.
         Ivanova, O. Melnikova,
         Int. J. Mod. Phys. E {\bf  5}, 191 (1996).

\bibitem{Blocki1977}
         J. Blocki, J. Randrup, W. J. Swiatecki, and C. F. Tsang,
         Ann. Phys. {\bf 105}, 427 (1977).


\bibitem{Proximity}
        G.L. Zhang, Y.J. Yao, M.F. Guo, M. Pan, G.X. Zhang, X.X. Liu,
        Nucl. Phys. A {\bf 951},  86 (2016).

\bibitem{Zag2006} V. I. Zagrebaev, Yu. Ts. Oganessian, M. G. Itkis, and W. Greiner, Phys. Rev C {\bf 73}, 031602(R) (2006).



\bibitem{SGZhou_DHbarrier}
       J. Zhao, B.-N. Lu, D. Vretenar, E.-G. Zhao, and S.-G. Zhou,
       Phys. Rev. C {\bf 91}, 014321 (2015).

\bibitem{Triple_Barr}
        J. Blons, C. Mazur, D. Paya, M. Ribrag, H. Weigmann,
        Nucl. Phys. A {\bf  414}, 1-41, (1984).

\bibitem{Kraszhoharkai}
       A. Krasznahorkay, D. Habs, M. Hunyadi, et al.,
       Phys. Lett. B {\bf 461}, 15 (1999).

\bibitem{Thirolf}
         L. Csige, M. Csatl\'{os}, T. Faestermann, et al.,
         Phys. Rev. C {\bf 80}, 011301(R) (2009).

\bibitem{Kraszhoharkai1}
        A. Krasznahorkay, M. Hunyadi, M. N. Harakeh, {\it et al.},
         Phys. Rev. Lett. {\bf 80}, 2073 (1998).

\bibitem{Simmons1965}
       J.E. Simmons, R. L. Henkel, Phys. Rev.{\bf 120}, 198 (1965);
        R.B. Leachman, L. Blumberg, Phys. Rev. {\bf 137}, B814 (1965).

\bibitem{huizenga}
         R. Vandenbosch, J.R. Huizenga,
         {\it Nuclear Fission}, Academic Press, New York and London, 1973.

\bibitem{bir}
         J.L. Birman, R.G. Nazmitdinov, V.I. Yukalov,
         Phys. Rep. {\bf 526}, 1 (2013).

\bibitem{Wong} C. Y. Wong, Phys. Rev. Lett.  {\bf 31}, 766 (1973).

\bibitem{andr}
        A.A. Andronov, A.A. Vitt, S.E. Khaikin,
        {\it Theory of Oscillators},
        International series of monographs in physics, v. 4,
        United Kingdom, Pergamon Press, 1966.

\bibitem{hopf}
        J. Marsden, M. McCracken,
        {\it Hopf Bifurcation and its Applications},
        New York Inc, Springer-Verlag, 1976.







\end{thebibliography}
\end{document}